\documentclass[prd,aps,nofootinbib,onecolumn,showpacs,12pt]{revtex4-1}
\usepackage{graphicx,epsfig}

\newcommand{\nn}{\nonumber}
\def\rar{\rightarrow}
\def\s1{\hat s}
\def\lb{\Lambda_b}
\def\R1{\varepsilon_1}
\def\E8{\varepsilon_8}

\def\ga{\gamma}

\newcommand{\f}{\frac}
\newcommand{\al}{\alpha_s}

\def\ds{\displaystyle}
\def\beq{\begin{equation}}
\def\eeq{\end{equation}}
\def\bea{\begin{eqnarray}}
\def\eea{\end{eqnarray}}
\def\beeq{\begin{eqnarray}}
\def\eeeq{\end{eqnarray}}

\def\vel{\left|}
\def\ver{\right|}
\def\nnb{\nonumber}
\def\ga{\left(}
\def\dr{\right)}

\def\rar{\rightarrow}
\def\nnb{\nonumber}

\def\ba{\begin{array}}
\def\ea{\end{array}}

\def\xis0{{\Xi^{*0}}}

\def\g5{\gamma_5}

\def\cp{&\times& \!\!\!}

\def\vel{\left|}
\def\ver{\right|}
\def\nnb{\nonumber}
\def\ga{\left(}
\def\dr{\right)}

\def\rar{\rightarrow}
\def\nnb{\nonumber}

\def\ba{\begin{array}}
\def\ea{\end{array}}

\def\xis0{{\Xi^{*0}}}

\def\g5{\gamma_5}

\def\es{\!\!\! & = & \!\!\!}

\def\ar{&+&~  \!\!\!}
\def\ek{&-&~ \!\!\!}

\def\cp{&\times& \!\!\!}

\usepackage{epsf}
\usepackage{graphicx}
\def\ds{\displaystyle}

\begin{document}
\title{ Analysis of $\Lambda_b \rar \Lambda \ell^+ \ell^-$ transition in SM4 using form factors from  Full QCD }
\author{  K. Azizi$^{\ddag}, $N. Kat{\i}rc{\i}$^{\dag}$\\
 Physics Department, Do\u gu\c s University, Ac{\i}badem-Kad{\i}k\"oy, \\ 34722 Istanbul, Turkey\\
$^\ddag$e-mail:kazizi@dogus.edu.tr\\
$^\dag$e-mail:nkatirci@dogus.edu.tr}

\begin{abstract}
Using the responsible form factors calculated via full QCD, we analyze the  $\Lambda_{b}\rightarrow \Lambda \ell^{+}\ell^{-}$ transition in
the standard model containing fourth  generation
quarks (SM4). We discuss  effects of the
presence of $t'$ fourth family quark on related observables like branching ratio,
forward-backward asymmetry, baryon polarization as well as double lepton polarization asymmetries. We also compare our results with those obtained in the  SM as well as with predictions of the SM4 but using form factors calculated
within heavy quark effective theory. The obtained results on branching ratio indicate  that the
$\Lambda_{b}\rightarrow \Lambda \ell^{+}\ell^{-}$ transition
 is more probable in full QCD comparing to the heavy quark effective theory. It is also shown that the results on all considered observables in SM4  deviate considerably from the SM predictions 
when $m_{t'}\geq 400~GeV$.

\end{abstract}

\maketitle
\section{Introduction}
The standard model (SM) has been the pillar of particle physics for
many years. However, there are some unsolved problems such as the origin of mass, the strong CP problem, neutrino oscillations,  origins of dark matter and dark energy,
 number of generations, matter-antimatter asymmetry, quantum gravity, unification and so on which can not be explained
by the SM. To cure such deficiencies, there exist various extensions of the standard model through supersymmetry, SM with fourth generation, etc. or entirely novel explanations,
 such as string theory, M-theory and extra dimensions.

The new theories beyond the SM need to be confirmed in the experiments. Hence, calculation of many parameters related to the decays of hadrons via new theories such as SM4 are important as  they  could be studied
at particle colliders. It is expected that the LHC will provide possibility to study properties of hadrons as well as
their electromagnetic, weak and strong decays. Among these decays, the weak decays of hadrons can play a crucial role in searching for physics beyond the SM. The loop level semileptonic weak
transitions of the heavy baryons containing single heavy quark to light baryons induced by
the flavor changing neutral currents (FCNC) are useful tools in this respect. In this connection, we analyze the  $\Lambda_b \rar \Lambda \ell^+ \ell^-$ transition in SM4 by calculating various related parameters
 like  branching ratio,
forward-backward asymmetry, baryon polarization as well as double lepton polarization asymmetries. Here, we use all involved twelve form factors recently calculated in full QCD \cite{kazizi}.
This work is an extension of the previous works \cite{kavb,vb,turan} where the two form factors  calculated within heavy quark effective theory (HQET) are used.

In the SM, the $\Lambda_b \rar \Lambda \ell^+ \ell^-$ channel proceeds via FCNC transition of $b \rar s \ell^+ \ell^-$ at quark level. The latter is described via a low energy effective Hamiltonian containing
 Wilson coefficients. In SM4, the form of Hamiltonian does not change but due to additional interactions of the fourth family quark $t'$ with other particles the Wilson coefficients are modified. Hence,
\bea\label{c4} C_7^{eff,tot}(m_{t^\prime},r_{sb},\phi_{sb}) &=&
C_7^{eff,SM} + \frac{\lambda_{t'}(r_{sb},\phi_{sb})}
{\lambda_t} C_7^{eff,new} (m_{t^\prime})~, \nnb \\
C_9^{eff,tot}(m_{t^\prime},r_{sb},\phi_{sb}) &=& C_9^{eff,SM} +
\frac{\lambda_{t'}(r_{sb},\phi_{sb})}
{\lambda_t}C_9^{eff,new} (m_{t^\prime}) ~, \nnb \\
C_{10}^{tot}(m_{t^\prime},r_{sb},\phi_{sb}) &=& C_{10}^{SM} +
\frac{\lambda_{t'}(r_{sb},\phi_{sb})} {\lambda_t} C_{10}^{new}
(m_{t^\prime})~, \eea where \bea
{\label{parameter}} \lambda_{t}&=&V_{t b} V_{t s}^\ast
~~~~~~\mbox{and} ~~~~~~~ \lambda_{t'}(r_{sb},\phi_{sb})=V_{t^\prime b} V_{t^\prime
s}^\ast=r_{sb}e^{i\phi_{sb}}.\eea  Here, $V_{t b}$, $ V_{t s}$ are
elements of Cabibbo-Kobayashi-Maskawa (CKM) matrix in the SM and
$V_{t' b}$, $ V_{t' s}$ are elements of the CKM matrix in the  SM4. In
the above relations, ($m_{t'}$, $r_{sb},\,\ \phi_{sb}$ ) is a set
of fourth generation parameters which we are going to discuss the
sensitivity of physical observables to them.  The new Wilson
coefficients,
 $C_7^{eff,new} (m_{t^\prime}),C_9^{eff,new}
(m_{t^\prime})$ and $C_{10}^{new} (m_{t^\prime})$ in Eqs.
(\ref{c4}) are obtained by replacing the mass of top quark by its
SM4 version ( $m_t \rar m_{t^\prime}$) \cite{R23,R25}.

It is expected that the masses of the fourth generation quarks are in the
interval (400-600) GeV \cite{soniyeni}.  As the mass difference between these two quarks is small,
we will refer to both members of the fourth family by $t'$. For the recent status of the SM4 quarks see for instance \cite{Holdom,Eberhardt,Sahin} and references therein.

The outline of the paper is as follows. In next section, we
present the effective Hamiltonian and transition matrix elements describing the 
decay under consideration. In  section III,
we  present the explicit expressions for physical observables
such as differential decay rate, forward backward asymmetry,
baryon polarization and double lepton polarization asymmetries.
This section also encompass  our numerical analysis on the  physical quantities under study as well as
our discussions. Finally, we will have a concluding section.

\section{The $\Lambda_b \rightarrow \Lambda \ell^+
\ell^-$ Transition}

\subsection{The Effective Hamiltonian}
 The quark structures of the initial and final baryons in  $\Lambda_b \rightarrow \Lambda \ell^+
\ell^-$ indicate that this channel proceeds via FCNC transition of $b \rightarrow s \ell^+\ell^-$, whose effective Hamiltonian in the SM is written as
\bea \label{e8401} {\cal H}^{eff} &=& {G_F \alpha_{em} V_{tb}
V_{ts}^\ast \over 2\sqrt{2} \pi} \Bigg[ C_9^{eff}
\bar{s}\gamma_\mu (1-\gamma_5) b \, \bar{\ell} \gamma^\mu \ell +
C_{10} \bar{s} \gamma_\mu (1-\gamma_5) b \, \bar{\ell} \gamma^\mu
\gamma_5 \ell \nnb \\
&-&  2 m_b C_7^{eff}{1\over q^2} \bar{s} i \sigma_{\mu\nu}q^{\nu}
(1+\gamma_5) b \, \bar{\ell} \gamma^\mu \ell \Bigg]~, \eea
where $G_F$ is the Fermi constant, $\alpha_{em}$  is the fine structure constant at Z
mass scale, and as we previously mentioned the $C_7^{eff}$, $C_9^{eff}$ and $C_{10}$ are the
Wilson coefficients representing different interactions. In the following, we present the explicit expressions of the Wilson coefficients in the SM. To get their expressions in SM4, it is enough  to apply 
Eq. (\ref{c4}). 

The $C_7^{eff}$ is given as \cite{AJBuras,R7627,R23}
\bea \label{e7603}
C_7^{eff}=\eta^{\frac{16}{23}} C_7(\mu_W)+\frac{8}{3} \left(
\eta^{\frac{14}{23}} -\eta^{\frac{16}{23}} \right)C_8(\mu_W)+C_2 (\mu_W) \sum_{i=1}^8 h_i \eta^{a_i}~, \nnb\\
\eea where
 \bea \eta =
\frac{\alpha_s(\mu_W)} {\alpha_s(\mu_b)} ~,~~~~ \mbox{and}~~~~~ \alpha_s(x)=
\frac{\alpha_s(m_Z)}{1-\beta_0\frac{\alpha_s(m_Z)}{2\pi}\ln(\frac{m_Z}{x})},
\eea
with $\alpha_s(m_Z)=0.118$ and $\beta_0=\frac{23}{3}$. The coefficients $a_i$ and $h_i$ are given as  \cite{R7627,R23}:
 \bea
   \label{klar}
\begin{array}{rrrrrrrrrl}
a_i = (\!\! & \f{14}{23}, & \f{16}{23}, & \f{6}{23}, & -
\f{12}{23}, &
0.4086, & -0.4230, & -0.8994, & 0.1456 & \!\!)  \vspace{0.1cm}, \\
h_i = (\!\! & 2.2996, & - 1.0880, & - \f{3}{7}, & - \f{1}{14}, &
-0.6494, & -0.0380, & -0.0186, & -0.0057 & \!\!).
\end{array}
\eea
The functions  $C_2(\mu_W)$, $C_7(\mu_W)$ and $C_8(\mu_W)$ inside the  $C_7^{eff}$ are
given as: \bea C_2(\mu_W)=1, ~~~ C_7(\mu_W)= -\frac{1}{2}
D_0(x_t)~,~~ C_8(\mu_W)= -\frac{1}{2} E_0(x_t)~ \eea
 where $x_t=\frac{m^2_t}{m^2_W}$ and 
\bea \label{e7604} D_0(x_t)= - \frac{(8 x_t^3+5 x_t^2-7 x_t)}{12
(1-x_t^3)}+ \frac{x_t^2(2-3 x_t)}{2(1-x_t)^4}\ln x_t~, \\ \nnb \\
\label{e7605} E_0(x_t)= - \frac{x_t(x_t^2-5x_t-2)}{4 (1-x_t^3)} +
\frac{3 x_t^2}{2 (1-x_t)^4}\ln x_t~. \eea

 The Wilson coefficient $C_{10}$ is given by \bea \label{e7616} C_{10}= -
\frac{Y(x_t)}{\sin^2 \theta_W}~ \eea where   $\sin^2\theta_W = 0.23$ and        
  \bea \label{e7612} Y(x_t)= \frac{x_t}{8} \left[ \frac{x_t -4}{x_t
-1}+\frac{3 x_t}{(x_t-1)^2} \ln x_t \right]~.
 \eea

In leading log approximation, the  Wilson coefficient $C_9^{eff}(s')$ entering the effective Hamiltonian of the  channel  under consideration can be written 
 as \cite{R7627,R23}:
 \bea \label{C9eff}
C_9^{eff}(\hat{s}') & = & C_9 \eta(\hat s') + h(z, \hat s')\left(
3 C_1 + C_2 + 3 C_3 + C_4 + 3
C_5 + C_6 \right) \nonumber \\
& & - \frac{1}{2} h(1, \hat s') \left( 4 C_3 + 4 C_4 + 3
C_5 + C_6 \right) \nonumber \\
& & - \frac{1}{2} h(0, \hat s') \left( C_3 + 3 C_4 \right) +
\frac{2}{9} \left( 3 C_3 + C_4 + 3 C_5 + C_6 \right) \eea where
\bea \eta(\hat s') & = & 1 +\frac{\al(\mu_b)}{\pi}\, \omega(\hat
s'),\eea \bea \label{omega} \omega(\hat s') & = & - \frac{2}{9}
\pi^2 - \frac{4}{3}\mbox{Li}_2(\hat s') - \frac{2}{3}
\ln \hat s' \ln(1-\hat s') - \frac{5+4\hat s'}{3(1+2\hat s')}\ln(1-\hat s') - \nonumber \\
& &  \frac{2 \hat s' (1+\hat s') (1-2\hat s')}{3(1-\hat s')^2
(1+2\hat s')} \ln \hat s' + \frac{5+9\hat s'-6\hat s'^2}{6 (1-\hat
s') (1+2\hat s')}, \eea with $\hat{s}'=\frac{q^2}{m_b^2}$. The allowed region for the transferred momentum square, $q^2$ is
 $4m_l^2\leq q^2\leq(m_{\Lambda_b}-m_\Lambda)^2$. The $C_9$ is
given as \bea C_9 = P_0^{NDR} +\frac{Y(x_t)}{\sin^2 \theta_W} - 4
Z(x_t),\eea where $P_0^{NDR}=2.60 \pm 0.25$ \cite{R7627,R23} in the naive dimensional
regularization scheme.

 The function, $Z(x_t)$ is defined  as:
 %\bea\label{e761} X(x_t)=
%\frac{x_t}{8} \left[ \frac{x_t+2}{x_t -1}+\frac{3
%x_t-6}{(x_t-1)^2} \ln x_t \right]~^,
% \eea
\bea \label{e7612} Z(x_t)=\frac{18 (x_t)^4-163 x_t^3+259 x_t^2
-108 x_t}{144 (x_t-1)^3} \nnb \left[\frac{32 (x_t)^4-38 x_t^3-15
x_t^2+18 x_t}{72
(x_t-1)^4} - \frac{1}{9}\right] \ln x_t. \\ \nnb \\
\eea
 The remaining coefficients in Eq. (\ref{C9eff}) is defined as:
 \bea \label{coeffs} C_j=\sum_{i=1}^8 k_{ji}
\eta^{a_i} \qquad (j=1,...6) \vspace{0.2cm} \eea where $k_{ji}$
 are given as:
\bea\frac{}{}
   \label{klar}
\begin{array}{rrrrrrrrrl}
k_{1i} = (\!\! & 0, & 0, & \f{1}{2}, & - \f{1}{2}, &
0, & 0, & 0, & 0 & \!\!)  \vspace{0.1cm}, \\
k_{2i} = (\!\! & 0, & 0, & \f{1}{2}, &  \f{1}{2}, &
0, & 0, & 0, & 0 & \!\!)  \vspace{0.1cm}, \\
k_{3i} = (\!\! & 0, & 0, & - \f{1}{14}, &  \f{1}{6}, &
0.0510, & - 0.1403, & - 0.0113, & 0.0054 & \!\!) , \vspace{0.1cm} \\
k_{4i} = (\!\! & 0, & 0, & - \f{1}{14}, &  - \f{1}{6}, &
0.0984, & 0.1214, & 0.0156, & 0.0026 & \!\!),  \vspace{0.1cm} \\
k_{5i} = (\!\! & 0, & 0, & 0, &  0, &
- 0.0397, & 0.0117, & - 0.0025, & 0.0304 & \!\!),  \vspace{0.1cm} \\
k_{6i} = (\!\! & 0, & 0, & 0, &  0, &
0.0335, & 0.0239, & - 0.0462, & -0.0112 & \!\!).  \vspace{0.1cm} \\
\end{array}
\eea
 Finally, the $h(y,
\hat s')$ function has the following explicit expression:
\bea \label{phasespace} h(y,
\hat s') & = & -\f{8}{9}\ln\f{m_b}{\mu_b} - \f{8}{9}\ln y +
\f{8}{27} + \f{4}{9} x \\
& & - \f{2}{9} (2+x) |1-x|^{1/2} \left\{
\begin{array}{ll}
\left( \ln\left| \f{\sqrt{1-x} + 1}{\sqrt{1-x} - 1}\right| - i\pi
\right), &
\mbox{for } x \equiv \f{4z^2}{\hat s'} < 1 \nonumber \\
2 \arctan \f{1}{\sqrt{x-1}}, & \mbox{for } x \equiv \f {4z^2}{\hat
s'} > 1,
\end{array}
\right. \\
\eea where   $y=1$ or $y=z=\frac{m_c}{m_b}$ and, \bea h(0, \hat
s') & = & \f{8}{27} -\f{8}{9} \ln\f{m_b}{\mu_b} - \f{4}{9} \ln
\hat s' + \f{4}{9} i\pi.\eea

\subsection{Transition Matrix Elements and Form Factors}
 The transition matrix elements for $\Lambda_b \rightarrow \Lambda
\ell^+ \ell^-$ are obtained by sandwiching  the
effective Hamiltonian between the initial and final baryonic
states. These matrix elements are parametrized in terms of twelve form factors in full QCD in the following way:
\bea\label{matrixel1a} \langle \Lambda(p) \mid  \bar s \gamma_\mu
(1-\gamma_5) b \mid \Lambda_b(p+q)\rangle = \bar {u}_\Lambda(p)
\Big[\gamma_{\mu}f_{1}(q^{2})+{i}
\sigma_{\mu\nu}q^{\nu}f_{2}(q^{2}) + q^{\mu}f_{3}(q^{2}) \nnb \\
-\gamma_{\mu}\gamma_5
g_{1}(q^{2})-{i}\sigma_{\mu\nu}\gamma_5q^{\nu}g_{2}(q^{2}) -
q^{\mu}\gamma_5 g_{3}(q^{2})
\vphantom{\int_0^{x_2}}\Big] u_{\Lambda_{b}}(p+q)~,\nnb \\
\eea \bea\label{matrixel1b} \langle \Lambda(p)\mid \bar s i
\sigma_{\mu\nu}q^{\nu} (1+ \gamma_5) b \mid
\Lambda_b(p+q)\rangle=\bar{u}_\Lambda(p)
\Big[\gamma_{\mu}f_{1}^{T}(q^{2})+{i}\sigma_{\mu\nu}q^{\nu}f_{2}^{T}(q^{2})+
q^{\mu}f_{3}^{T}(q^{2}) \nnb \\
+ \gamma_{\mu}\gamma_5
g_{1}^{T}(q^{2})+{i}\sigma_{\mu\nu}\gamma_5q^{\nu}g_{2}^{T}(q^{2})
+ q^{\mu}\gamma_5 g_{3}^{T}(q^{2})
\vphantom{\int_0^{x_2}}\Big] u_{\Lambda_{b}}(p+q)~,\nnb \\
\eea
where $f_{1}$, $f_{2}$, $f_{3}$,
$g_{1}$, $g_{2}$, $g_{3}$, $f^T_{1}$, $f^T_{2}$, $f^T_{3}$, $g^T_{1}$, $g^T_{2}$ and
$g^T_{3}$ are transition form factors in full theory. These form factors have been recently calculated in \cite{kazizi} in the framework of light cone QCD sum rules. 

In the HQET, the twelve form factors in full QCD reduce to two form
factors, $F_1$ and $F_2$, hence the transition matrix element in this limit is  defined as
\cite{alievozpineci,Mannel}:
 \bea\label{matrixel1111} \langle
\Lambda(p) \mid \bar s\Gamma b\mid \Lambda_b(p+q)\rangle=
\bar{u}_\Lambda(p)[F_1(q^2)+\not\!vF_2(q^2)]\Gamma
u_{\Lambda_b}(p+q), \eea where $\Gamma$ denotes   any Dirac
matrices and $\not\!v=({\not\!p}+{\not\!q})/m_{\Lambda_{b}}$.
These  form factor are calculated in \cite{huang}. 
Comparing the definitions of the transition matrix
elements both in full QCD and HQET theories, one can easily find the following relations among the above mentioned form
factors: 
\bea\label{matrixel22222}
 f_{1}&= &g_{1}=f_{2}^{T} = g_{2}^{T}=F_1+ \frac{m_\Lambda}{m_{\Lambda_b}}
F_2~,\nnb\\
f_2 &= &g_2 = f_3 = g_3=\frac{F_2}{m_{\Lambda_b}}~,\nnb\\
f_1^{T} &=& g_1^{T} =\frac{F_2}{m_{\Lambda_b}}q^2~,\nnb\\
f_3^{T} &=&-\frac{F_2}{m_{\Lambda_b} }(m_{\Lambda_b}-m_{\Lambda})~,\nnb\\
g_3^{T} &=& \frac{F_2}{m_{\Lambda_b} }(m_{\Lambda_b}+m_{\Lambda})~.
\eea

\section{Physical Observables characterizing the  $\Lambda_b \rightarrow \Lambda
\ell^+ \ell^-$ transition }
\subsection{Branching Ratio}
Using the decay amplitude and transition matrix elements in terms of form factors, the differential decay rate is obtained as a function of SM4 parameters  as
\cite{R7601,savcibey,Giri}:
 \small{ \bea\frac{d\Gamma}{d\hat s dz}(z,\hat
s,m_{t^\prime},r_{sb},\phi_{sb}) = \frac{G_F^2\alpha^2_{em}
m_{\Lambda_b}}{16382 \pi^5}| V_{tb}V_{ts}^*|^2 v \sqrt{\lambda}
\Bigg[{\cal T}_0(\hat
s,m_{t^\prime},r_{sb},\phi_{sb}) \nnb\\
+{\cal T}_1(\hat s,m_{t^\prime},r_{sb},\phi_{sb}) z+{\cal
T}_2(\hat s,m_{t^\prime},r_{sb},\phi_{sb}) z^2\Bigg]~,
\label{rate} \eea}
 where $z=\cos\theta$ with
 $\theta$ being the angle between the momenta of 
$\Lambda_b$ and $\ell^-$ in the center of mass of leptons,
$\lambda=\lambda(1,r,\hat s)=1+r^2+\hat s^2-2r-2\hat s-2r\hat s$,
$r= m^2_{\Lambda}/m^2_{\Lambda_b}$ and $v=\sqrt{1-\frac{4
m_\ell^2}{q^2}}$. Here,  $\hat s=\frac{q^2}{m_{\Lambda_b}^2}$ and  we have the relation, $ \hat s'=\frac{\hat s m_{\Lambda_b}^2}{m_b^2}$ between the $\hat s$ and previously used  $ \hat s'$. The functions, ${\cal T}_0(\hat
s,m_{t^\prime},r_{sb},\phi_{sb})$,  ${\cal
T}_1(\hat
s,m_{t^\prime},r_{sb},\phi_{sb})$ and ${\cal T}_2(\hat
s,m_{t^\prime},r_{sb},\phi_{sb})$ are given as ( see also
\cite{kazizi}):
 \bea {\cal T}_0(\hat
s,m_{t^\prime},r_{sb},\phi_{sb}) &=& 32 m_\ell^2
m_{\Lambda_b}^4 \hat s (1+r-\hat s) \ga \vel D_3 \ver^2 +
\vel E_3 \ver^2 \dr \nnb \\
\ar 64 m_\ell^2 m_{\Lambda_b}^3 (1-r-\hat s) \, \mbox{\rm Re}
[D_1^\ast E_3 + D_3
E_1^\ast] \nnb \\
\ar 64 m_{\Lambda_b}^2 \sqrt{r} (6 m_\ell^2 - m_{\Lambda_b}^2 \hat
s)
{\rm Re} [D_1^\ast E_1] \nnb \\
\ar 64 m_\ell^2 m_{\Lambda_b}^3 \sqrt{r} \Big( 2 m_{\Lambda_b}
\hat s {\rm Re} [D_3^\ast E_3] + (1 - r + \hat s)
{\rm Re} [D_1^\ast D_3 + E_1^\ast E_3]\Big) \nnb \\
\ar 32 m_{\Lambda_b}^2 (2 m_\ell^2 + m_{\Lambda_b}^2 \hat s)
\Big\{ (1 - r + \hat s) m_{\Lambda_b} \sqrt{r} \,
\mbox{\rm Re} [A_1^\ast A_2 + B_1^\ast B_2] \nnb \\
\ek m_{\Lambda_b} (1 - r - \hat s) \, \mbox{\rm Re} [A_1^\ast B_2
+ A_2^\ast B_1] - 2 \sqrt{r} \Big( \mbox{\rm Re} [A_1^\ast B_1] +
m_{\Lambda_b}^2 \hat s \,
\mbox{\rm Re} [A_2^\ast B_2] \Big) \Big\} \nnb \\
\ar 8 m_{\Lambda_b}^2 \Big\{ 4 m_\ell^2 (1 + r - \hat s) +
m_{\Lambda_b}^2 \Big[(1-r)^2 - \hat s^2 \Big]
\Big\} \ga \vel A_1 \ver^2 +  \vel B_1 \ver^2 \dr \nnb \\
\ar 8 m_{\Lambda_b}^4 \Big\{ 4 m_\ell^2 \Big[ \lambda + (1 + r -
\hat s) \hat s \Big] + m_{\Lambda_b}^2 \hat s \Big[(1-r)^2 - \hat
s^2 \Big]
\Big\} \ga \vel A_2 \ver^2 +  \vel B_2 \ver^2 \dr \nnb \\
\ek 8 m_{\Lambda_b}^2 \Big\{ 4 m_\ell^2 (1 + r - \hat s) -
m_{\Lambda_b}^2 \Big[(1-r)^2 - \hat s^2 \Big]
\Big\} \ga \vel D_1 \ver^2 +  \vel E_1 \ver^2 \dr \nnb \\
\ar 8 m_{\Lambda_b}^5 \hat s v^2 \Big\{ - 8 m_{\Lambda_b} \hat s
\sqrt{r}\, \mbox{\rm Re} [D_2^\ast E_2] +
4 (1 - r + \hat s) \sqrt{r} \, \mbox{\rm Re}[D_1^\ast D_2+E_1^\ast E_2]\nnb \\
\ek 4 (1 - r - \hat s) \, \mbox{\rm Re}[D_1^\ast E_2+D_2^\ast E_1]
+ m_{\Lambda_b} \Big[(1-r)^2 -\hat s^2\Big] \ga \vel D_2 \ver^2 +
\vel
E_2 \ver^2\dr \Big\},\nnb \\
\eea \bea {\cal T}_1(\hat
s,m_{t^\prime},r_{sb},\phi_{sb}) &=& -16 m_{\lb}^4\s1 v \sqrt{\lambda}
\Big\{ 2 Re(A_1^* D_1)-2Re(B_1^* E_1)\nn\\
&+& 2m_{\lb}
Re(B_1^* D_2-B_2^* D_1+A_2^* E_1-A_1^*E_2)\Big\}\nn\\
&+&32 m_{\lb}^5 \s1~ v \sqrt{\lambda} \Big\{
m_{\lb} (1-r)Re(A_2^* D_2 -B_2^* E_2)\nn\\
&+& \sqrt{r} Re(A_2^* D_1+A_1^* D_2-B_2^*E_1-B_1^* E_2)\Big\}\;,
\eea
 \bea {\cal T}_2(\hat
s,m_{t^\prime},r_{sb},\phi_{sb}) &=& - 8 m_{\Lambda_b}^4 v^2 \lambda \ga
\vel A_1 \ver^2 + \vel B_1 \ver^2 + \vel D_1 \ver^2
+ \vel E_1 \ver^2 \dr \nnb \\
\ar 8 m_{\Lambda_b}^6 \hat s v^2 \lambda \Big( \vel A_2 \ver^2 +
\vel B_2 \ver^2 + \vel D_2 \ver^2 + \vel E_2 \ver^2  \Big)~, \eea
where,
 \bea \label{a9} A_1&=& A_1(\hat
s,m_{t^\prime},r_{sb},\phi_{sb})\nnb \\&=&\frac{1}{\hat sm_{\Lambda_b}^2}\ga f_1^T(\hat s)+g_1^T(\hat s) \dr \ga -2 m_b C_7^{eff}(\hat
s,m_{t^\prime},r_{sb},\phi_{sb})\dr + \Big( f_1(\hat s)-g_1(\hat s) \Big) C_9^{eff}(\hat
s,m_{t^\prime},r_{sb},\phi_{sb}) \nnb \\
A_2&=& A_1 \ga 1 \rar 2 \dr ~,\nnb \\
A_3&=& A_1 \ga 1 \rar 3 \dr ~,\nnb \\
B_1&=&A_1 \ga g_1(\hat s) \rar - g_1(\hat s);~g_1^T(\hat s) \rar - g_1^T(\hat s) \dr ~,\nnb \\
B_2&=&B_1 \ga 1 \rar 2 \dr ~,\nnb \\
B_3&=&B_1 \ga 1 \rar 3 \dr ~,\nnb \\
D_1&=&\Big( f_1(\hat s)-g_1(\hat s) \Big) C_{10}(\hat
s,m_{t^\prime},r_{sb},\phi_{sb})~,\nnb \\
D_2&=& D_1 \ga 1 \rar 2 \dr ~, \nnb \\
D_3&=&D_1 \ga 1 \rar 3 \dr ~,\nnb \\
E_1&=&D_1 \Big( g_1(\hat s) \rar - g_1(\hat s) \Big) ~,\nnb \\
E_2&=&E_1 \ga 1 \rar 2 \dr ~,\nnb \\
E_3&=&E_1 \ga 1 \rar 3 \dr ~.
 \eea
Integrating the aforementioned angular dependent differential decay rate over $z$, we get the $\hat s$ and SM4 parameters  dependent differential decay width  as
\bea\small{ \frac{d\Gamma(\hat s,m_{t^\prime},
r_{sb},\phi_{sb})}{d \hat s} = \frac{G_F^2\alpha^2_{em}
m_{\Lambda_b}}{8192 \pi^5}| V_{tb}V_{ts}^*|^2 v \sqrt{\lambda} \,
 \Delta(\hat s,m_{t^\prime},
r_{sb},\phi_{sb})~, \label{rate} }\eea
where, \bea \Delta(\hat s,m_{t^\prime},
r_{sb},\phi_{sb})={{\cal T}_0(\hat s,m_{t^\prime},
r_{sb},\phi_{sb}) +\frac{1}{3} {\cal T}_2(\hat s,m_{t^\prime},
r_{sb},\phi_{sb})}.\eea

Performing integration over  $\hat s$ in the kinematical region $\frac{4
m_\ell^2}{m_{\Lambda_b}^2}\le\hat s \le (1-\sqrt{r})^2$, the total decay width is obtained. Finally, using the lifetime of the $\Lambda_b$ baryon, we obtain  the branching
ratio depending on SM4 parameters. 

In further numerical analysis, we take the values, $m_t=167~GeV$, $m_W=80.4~GeV$,
$m_b=4.8~GeV$, $m_c=1.35~GeV$, $\mu_b=5~GeV$,
$\mu_W=80.4~GeV$, $m_e=0.00051$, $m_{\tau}=1.778$,
$m_{\mu}=0.105~GeV$, $| V_{tb}V_{ts}^\ast|=0.041$, $G_F = 1.166
\times 10^{-5}~ GeV^{-2}$, $\alpha_{em}=\frac{1}{129}$,
$\tau_{\Lambda_b}=1.383\times 10^{-12}~s$, $m_\Lambda =
1.116~GeV$ and $m_{\Lambda_{b}} = 5.624~ GeV$. The
present SM measurements and unitarity condition of the CKM matrix
imply that 
\cite{30,32,A. K. Alok,31} \bea \label{vtb} r_{sb}=\vel
V_{t^{\prime}b}V_{t^{\prime}s}^* \ver \leq 1.5 \times 10^{-2}.\eea
In our numerical calculations, we will consider the three different values $r_{sb}=\vel
V_{t^{\prime}b}V_{t^{\prime}s}^* \ver =0.005,~0.010$ and $0.015 $.
As we previously mentioned,  the masses of the fourth generation quarks are expected to be  in the
interval (400-600) GeV. In the present work, we will plot our figures considering the $m_{t'}$ in the interval (175-600) GeV to  see better at which points  the SM4 results start to deviate from the usual SM predictions.
 The  $\phi_{sb}$ is taken as $\phi_{sb}=\frac{\pi}{2}$  \cite{W. S.} (see also \cite{aslammm}).

The dependence of the branching ratio of the channel under consideration   for the $\mu$ and $\tau$ leptons on $m_{t'}$ at three fixed values of the  $r_{sb}$ as well as the SM are presented in figures 1 and 2. In these figures,  
the left graph corresponds to the HQET while the graph on the right refers to the full QCD.
We take into account the errors of the form factors in our analysis, hence we have a bound for each SM and SM4 with three different values of the $r_{sb}$ obtained from adding (subtracting) of the uncertainties 
to (from)
the central values. 
 
\begin{figure}[!h]
\label{fig2} \centering
\begin{tabular}{cc}
\epsfig{file=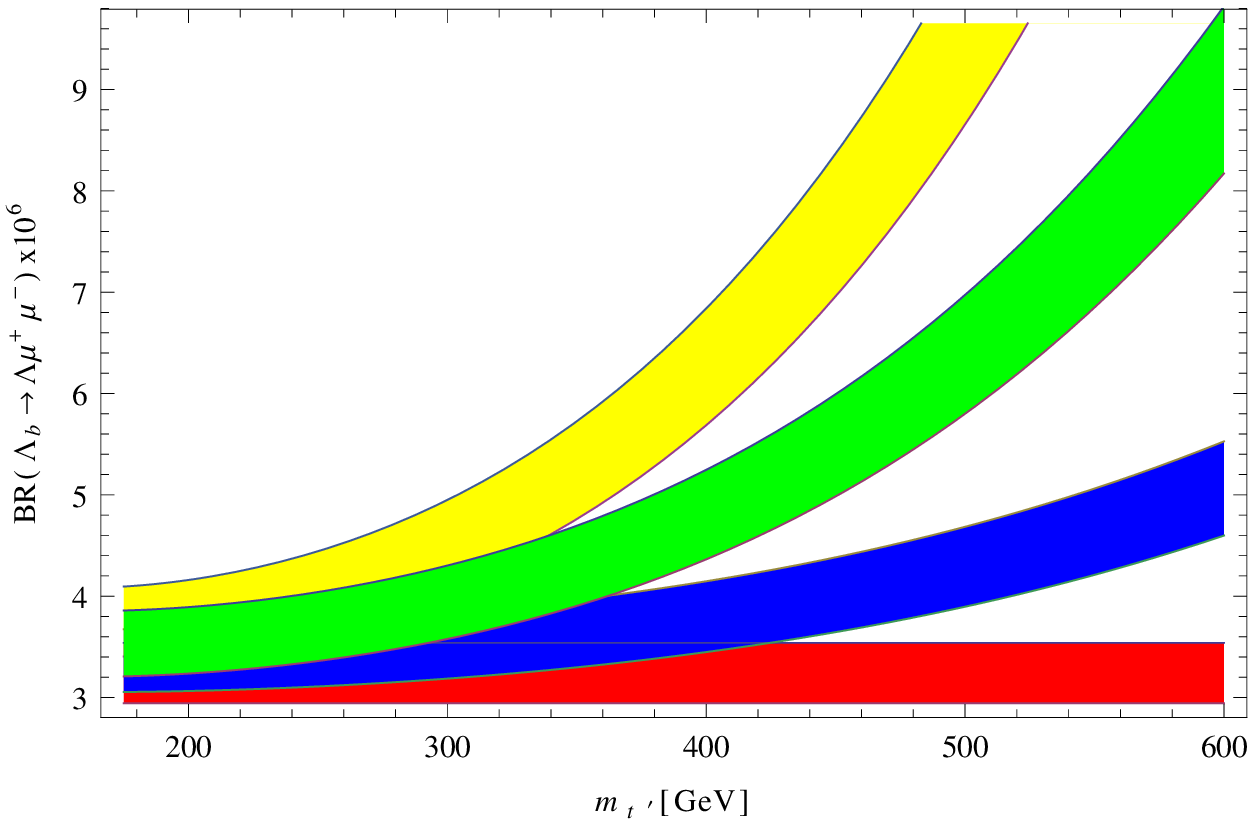,width=0.45\linewidth,clip=} &
\epsfig{file=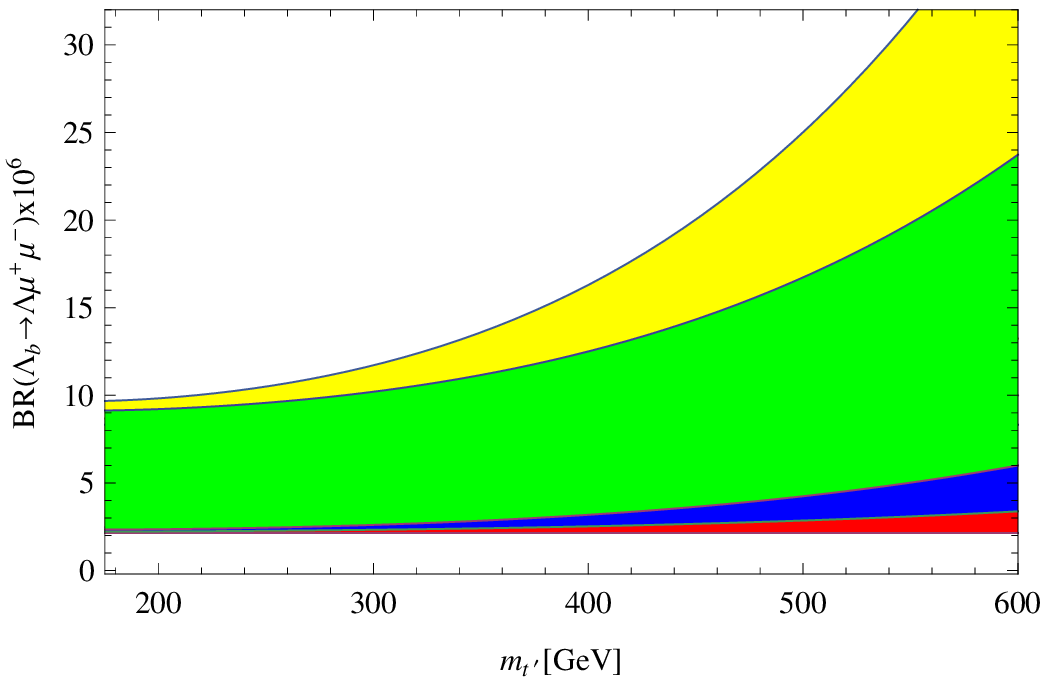,width=0.45\linewidth,clip=}
\end{tabular}
\caption{ The dependence of branching ratio for the $\Lambda_b
\rar \Lambda \mu^+ \mu^-$ decay on $m_{t'}$. The red  band corresponds to  the  SM, while  the blue, green and yellow bands belong 
to the SM4 for  $r_{sb} = 0.005, ~0.01$ and $0.015$, respectively. The left graph corresponds to the HQET while the graph on the right refers to the full QCD.}
\end{figure}
\begin{figure}[!h]
\label{fig1} \centering
\begin{tabular}{cc}
\epsfig{file=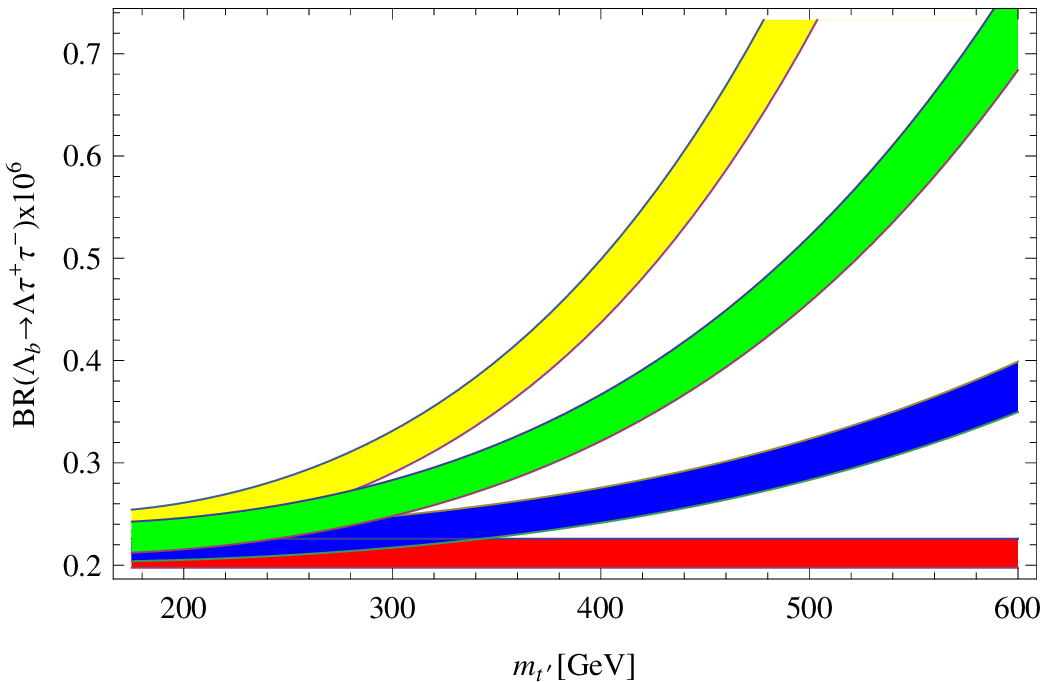,width=0.45\linewidth,clip=} &
\epsfig{file=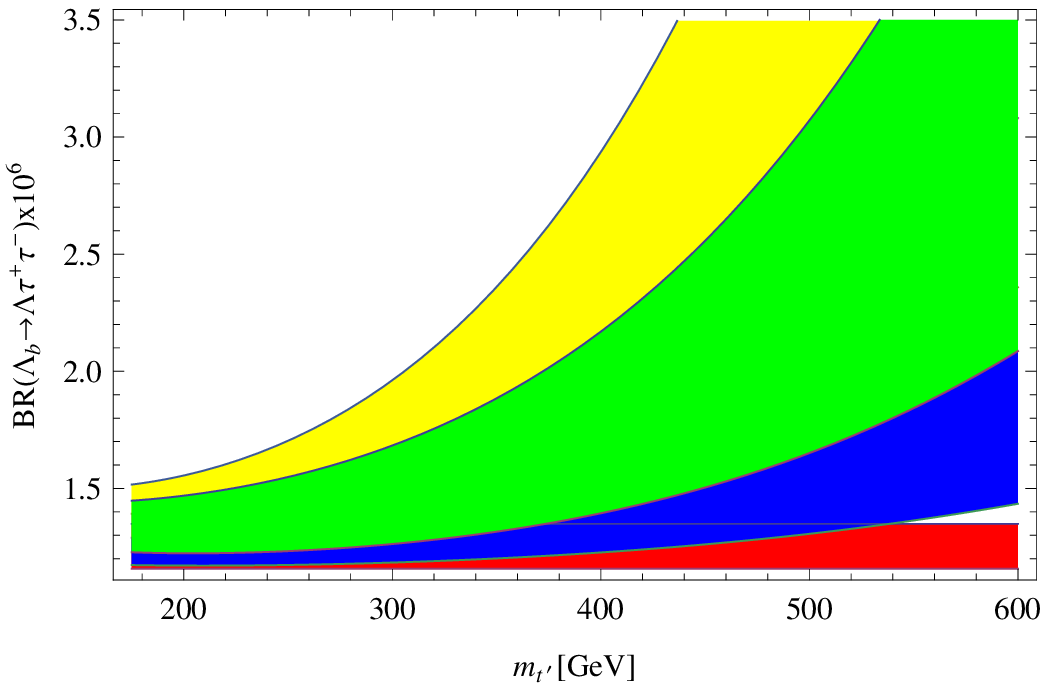,width=0.45\linewidth,clip=}
\end{tabular}
\caption{ The same as FIG. 1 but for $\tau$.}
\end{figure}
From   these figures, we see that
\begin{itemize}
\item in all cases, the branching ratios in SM4 grow increasing the fourth generation quark mass. The  deviation of the SM4 results from those of the SM becomes important at $m_{t'}\simeq 400~GeV$ and our results 
favor  $m_{t'}\geq 400~GeV$.
This is in good consistency with the results of \cite{soniyeni} in explanation of the observed CP asymmetries in the $B$ and $B_s$ decays.

\item Increasing in the $r_{sb}$ leads to an increase  in the value of the  branching ratio in all cases. The maximum deviation of the SM4 results from those of the SM belong to the $r_{sb}=0.015$ at 
any fixed values
of the $m_{t'}$ in the interval $400~GeV\leq m_{t'}\leq 600~GeV$. As far as the branching ratio is concerned, the difference between the SM and SM4 results with  $r_{sb}=0.005$ is considerable in HQET approximation but the uncertainties of 
the form factors approximately kill this difference
in full theory. For  $r_{sb}\in[0.1-0.15]$ the deviation of the SM4 results from those of the SM cannot be killed by the errors of the form factors in both HQET and Full theories. Such considerable discrepancy can be
considered as an indication for existing the fourth generation of the quarks.
 
\item As it is expected, the branching ratios in $\tau$ channel are small compared to the $\mu$ channel.
\end{itemize}
\begin{figure}[!h]
\label{fig1} \centering
\begin{tabular}{cc}
\epsfig{file=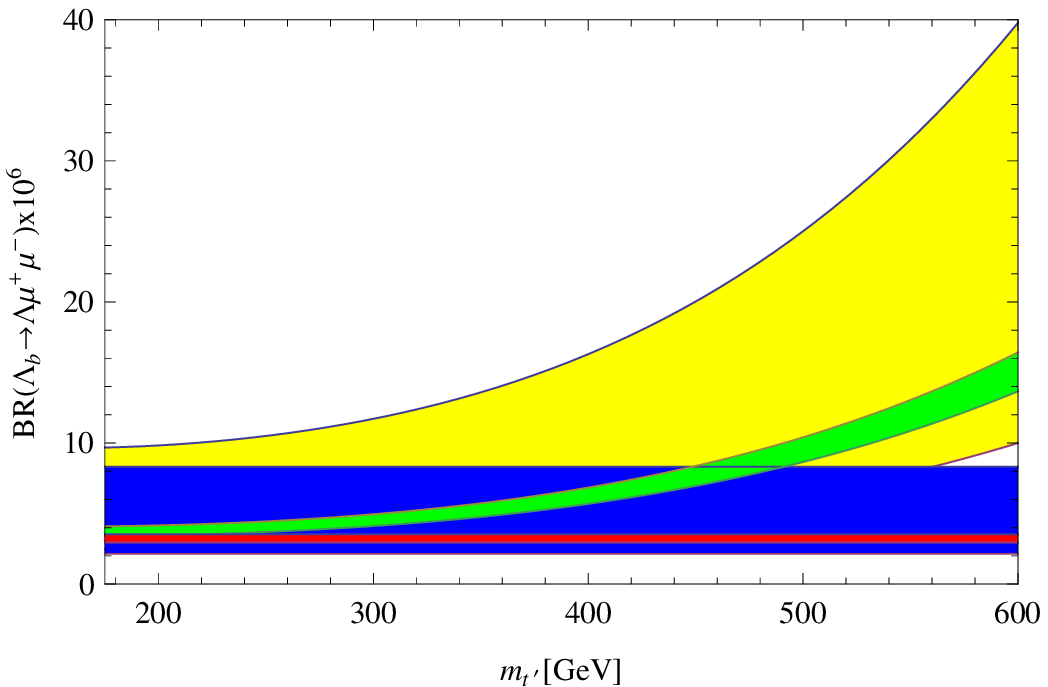,width=0.45\linewidth,clip=} &
\epsfig{file=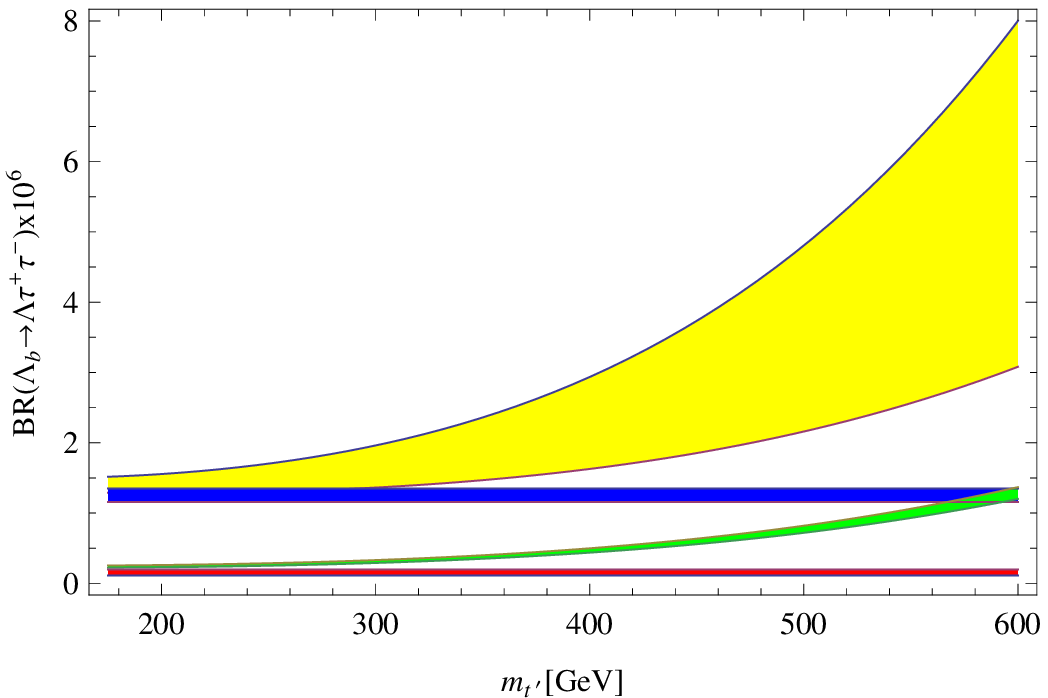,width=0.45\linewidth,clip=}
\end{tabular}
\caption{Comparison of the  branching ratio for the $\Lambda_b\rar \Lambda l^+ l^-$ decay in full QCD and HQET.  The blue and yellow bands respectively correspond to the SM and SM4 ($r_{sb} = 0.015$)  in full QCD,
 while the red and green bands respectively refer  to the SM and SM4 ($r_{sb} = 0.015$) in HQET.}
\end{figure}

We also  compare the full QCD and HQET results of the branching ratios obtained from the SM and SM4 with only $r_{sb} = 0.015$ together in figure 3 for both leptons. Looking at this figure, we deduce that
\begin{itemize}
\item  the full QCD results on branching ratios sweep large areas compared to those of the HQET. As far as the  branching ratios are considered, the SM and SM4 with $r_{sb} = 0.015$ bands obtained from the HQET
lie inside the bands of the full QCD in $\mu$ channel but we see considerable discrepancy between predictions of  these theories in the $\tau$ channel. 

\end{itemize}

\subsection{Forward-backward asymmetry}
The forward-backward asymmetry refers to the difference between the number of particles
that move on the forward  and those move on the backward  direction. It   is one of the promising tools in looking for new
physics beyond the SM. The SM4 parameters dependent forward-backward asymmetry is defined as:
\bea {\cal A}_{FB} (\hat s,m_{t^\prime}, r_{sb},\phi_{sb})=
\frac{\ds{\int_0^1\frac{d\Gamma}{d\hat{s}dz}}(z,\hat
s,m_{t^\prime}, r_{sb},\phi_{sb})\,dz -
\ds{\int_{-1}^0\frac{d\Gamma}{d\hat{s}dz}}(z,\hat s,m_{t^\prime},
r_{sb},\phi_{sb})\,dz}
{\ds{\int_0^1\frac{d\Gamma}{d\hat{s}dz}}(z,\hat s,m_{t^\prime},
r_{sb},\phi_{sb})\,dz +
\ds{\int_{-1}^0\frac{d\Gamma}{d\hat{s}dz}}(z,\hat s,m_{t^\prime},
r_{sb},\phi_{sb})\,dz}.~ \eea  
Using the $\hat s$, $z$ and fourth family parameters dependent differential decay rate we plot the ${\cal A}_{FB} $ in terms of $m_{t'}$ at $\hat s=0.5$ and at three fixed values of the $r_{sb}$
 and the SM in
 figures 4 and 5.
\begin{figure}[!h]
\label{fig4} \centering
\begin{tabular}{cc}
\epsfig{file=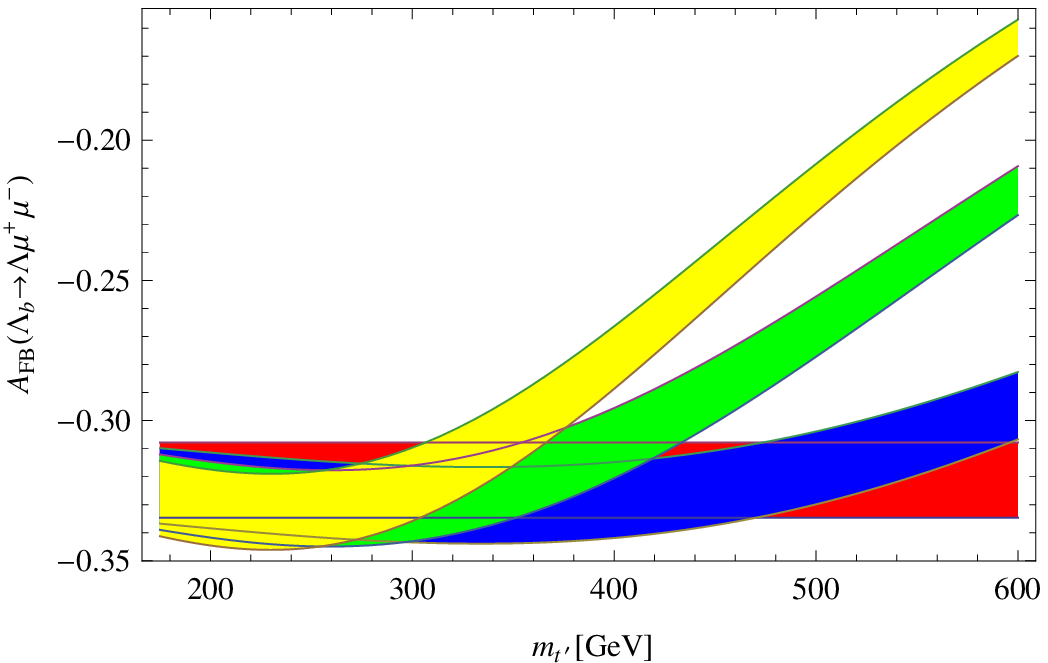,width=0.45\linewidth,clip=} &
\epsfig{file=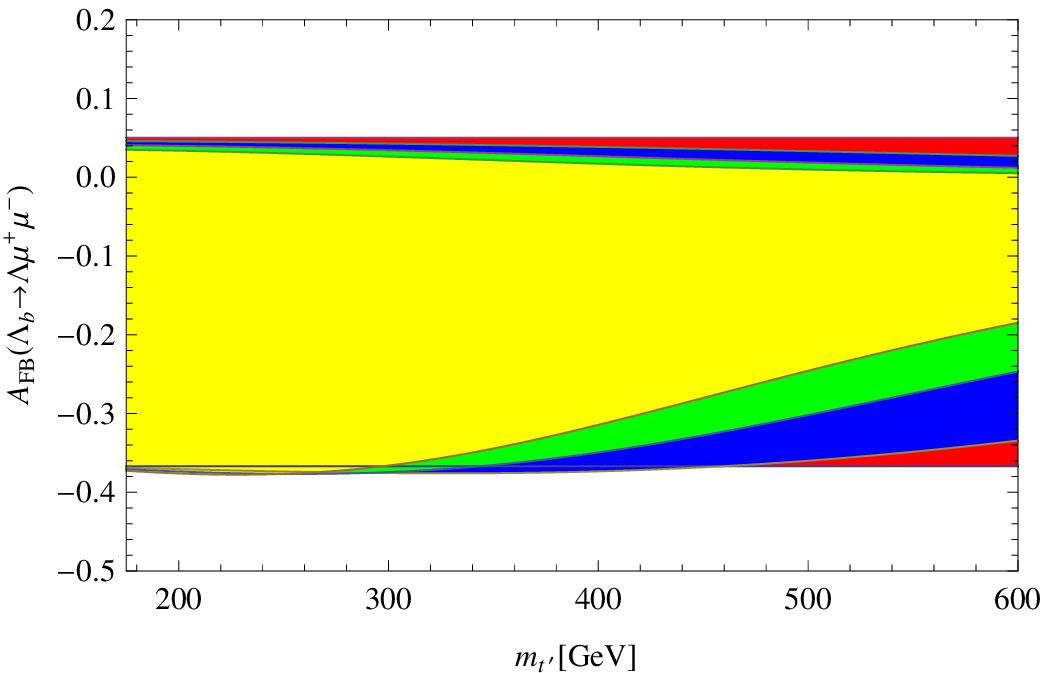,width=0.45\linewidth,clip=}
\end{tabular}
\caption{ The dependence of forward-backward asymmetry for the $\Lambda_b
\rar \Lambda \mu^+ \mu^-$ decay on $m_{t'}$ at $\hat s=0.5$. The red  band corresponds to  the  SM, while  the blue, green and yellow bands belong 
to the SM4 for  $r_{sb} = 0.005, ~0.01$ and $0.015$, respectively. The left graph corresponds to the HQET while the graph on the right refers to the full QCD.}
\end{figure}

\begin{figure}[!h]
\label{fig3} \centering
\begin{tabular}{cc}
\epsfig{file=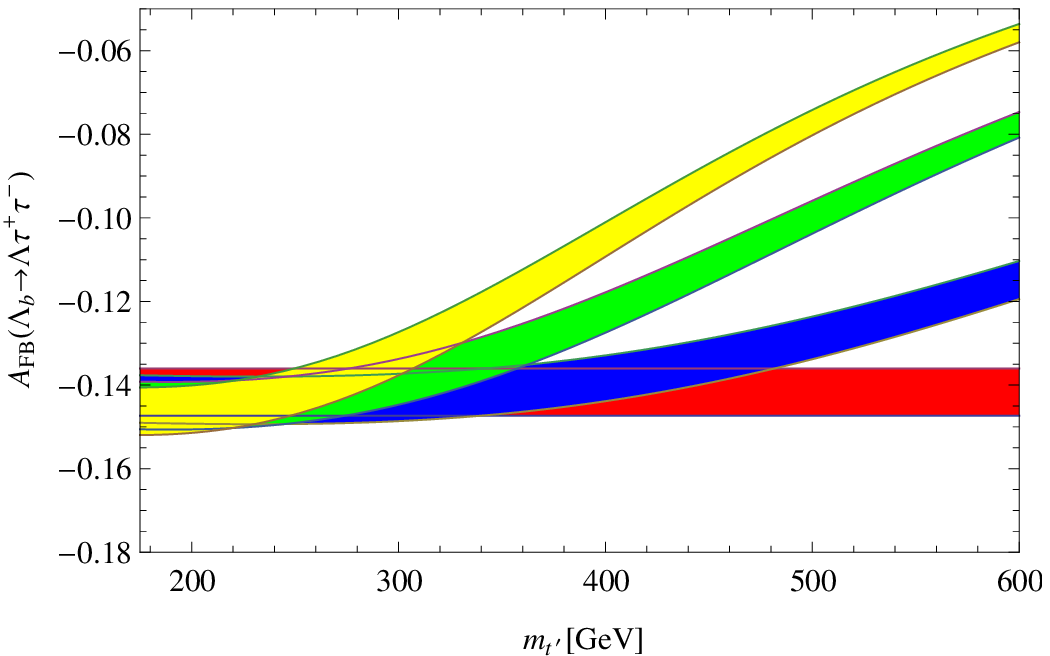,width=0.45\linewidth,clip=} &
\epsfig{file=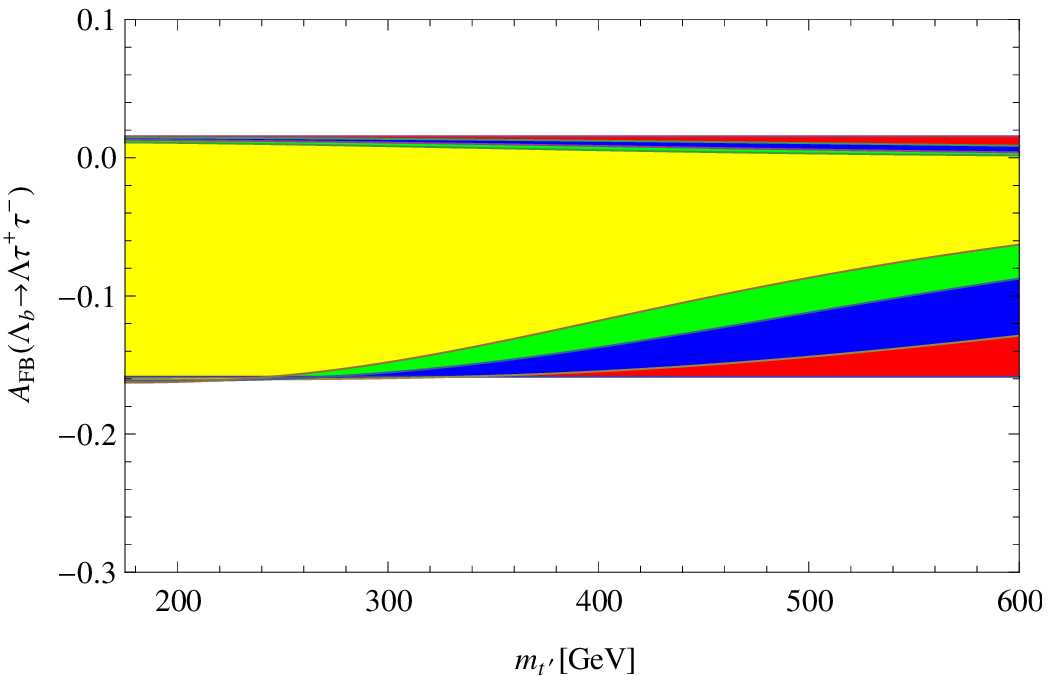,width=0.45\linewidth,clip=}
\end{tabular}
\caption{The same as FIG. 4 but for $\tau$. }
\end{figure}
From these figures, it is clear that, 
\begin{itemize}
\item our analysis on the forward-backward asymmetry also seems  to favor  $m_{t'}\geq 400~GeV$.

 \item There is considerable  HQET violations in both lepton channels. The difference between the predictions of the full theory and HQET is large in $\mu$ channel compared to that of the  $\tau$.

\item  There are considerable discrepancies between the SM4 and the SM results at high $m_{t'}$ values in HQET theory for both leptons. However, the uncertainties of the form factors in full theory suppress
 these differences such that the results of SM4 for all values of the $r_{sb}$ and $m_{t'}$ lie inside the SM bands.
\end{itemize}
\subsection{Baryon Polarizations} 
The definitions for the normal ($P_N$), longitudinal
($P_L$)  and transversal ($P_T$) polarizations of the $\Lambda$
baryon in the massive lepton case, are given in \cite{R7615}. Using those definitions, 
the general model independent expressions for the above polarizations are calculated in
\cite{30ozpineci,ozpinecibey}). In the case of SM4, those expressions reduce to the following explicit forms:
\bea \label{a2} P_N (\hat s,m_{t^\prime}, r_{sb},\phi_{sb})~\es~
\frac{8 \pi m_{\Lambda_b}^3 v \sqrt{\hat s}}{\Delta (\hat
s,m_{t^\prime}, r_{sb},\phi_{sb})}\Bigg\{ - 2 m_{\Lambda_b}
(1-r+\hat s) \sqrt{r} \,
\mbox{\rm Re}[A_1^\ast D_1 + B_1^\ast E_1] \nnb \\
\ar m_{\Lambda_b} (1-\sqrt{r}) [(1+\sqrt{r})^2 -\hat s] \, \Big(
 m_\ell \mbox{\rm Re}[(A_2-B_2)^\ast F_1] \Big) \nnb \\
\ar m_\ell [(1+\sqrt{r})^2 -\hat s] \,
\mbox{\rm Re}[A_1^\ast F_1] \nnb \\
\ar 4 m_{\Lambda_b}^2 \hat s \sqrt{r} \, \mbox{\rm Re}[A_1^\ast
E_2 + A_2^\ast E_1 +B_1^\ast D_2 +
B_2^\ast D_1] \nnb \\
\ek 2 m_{\Lambda_b}^3 \hat s \sqrt{r} (1-r+\hat s) \,
\mbox{\rm Re}[A_2^\ast D_2 + B_2^\ast E_2^\ast] \nnb \\
\ar 2 m_{\Lambda_b} (1-r-\hat s) \, \Big( \mbox{\rm Re}[A_1^\ast
E_1 + B_1^\ast D_1] +
m_{\Lambda_b}^2 \hat s \mbox{\rm Re}[A_2^\ast E_2 + B_2^\ast D_2] \Big) \nnb \\
\ek m_{\Lambda_b}^2 [(1-r)^2-\hat s^2] \, \mbox{\rm Re}[A_1^\ast
D_2 + A_2^\ast D_1 + B_1^\ast E_2 +
B_2^\ast E_1] \nnb \\
\ek m_\ell [(1+\sqrt{r})^2 -\hat s] \, \mbox{\rm Re}[B_1^\ast F_1]
\Bigg\}~ , \eea

\bea \label{a1} P_L (\hat s,m_{t^\prime}, r_{sb},\phi_{sb})~\es~
\frac{16 m_{\Lambda_b}^2 \sqrt{\lambda}}{\Delta(\hat
s,m_{t^\prime}, r_{sb},\phi_{sb})} \Bigg\{ 8 m_\ell^2
m_{\Lambda_b}\, \Big( \mbox{\rm Re}[D_1^\ast E_3 - D_3^\ast E_1] +
\sqrt{r} \mbox{\rm Re}[D_1^\ast D_3 - E_1^\ast E_3)] \Big) \nnb \\
\ar 2 m_\ell m_{\Lambda_b}\,  (1+\sqrt{r}) \mbox{\rm
Re}[(D_1-E_1)^\ast F_2]
 \nnb \\
\ek 2 m_\ell m_{\Lambda_b}^2 \hat s \, \Big\{ \mbox{\rm
Re}[(D_3-E_3)^\ast F_2 ] +
2 m_\ell ( \vel D_3 \ver^2 - \vel E_3 \ver^2 ) \Big\} \nnb \\
\ek 4 m_{\Lambda_b} (2 m_\ell^2 + m_{\Lambda_b}^2 \hat s) \,
\mbox{\rm Re}[A_1^\ast B_2 - A_2^\ast B_1] \nnb \\
\ek \frac{4}{3} m_{\Lambda_b}^3 \hat s v^2 \, \Big( 3 \mbox{\rm
Re}[D_1^\ast E_2 - D_2^\ast E_1] +
\sqrt{r} \mbox{\rm Re}[D_1^\ast D_2 - E_1^\ast E_2] \Big) \nnb \\
\ek \frac{4}{3} m_{\Lambda_b} \sqrt{r} (6 m_\ell^2 +
m_{\Lambda_b}^2 \hat s v^2) \, \mbox{\rm Re}[A_1^\ast A_2 - B_1^\ast B_2] \nnb \\
\ar \frac{1}{3} \Big\{ 3 [4 m_\ell^2 + m_{\Lambda_b}^2 (1-r+\hat
s)] (\vel A_1 \ver^2 -
\vel B_1 \ver^2 ) - 3 [4 m_\ell^2 -  m_{\Lambda_b}^2 (1-r+\hat s)] \nnb \\
\cp (\vel D_1 \ver^2 - \vel E_1 \ver^2 ) -  m_{\Lambda_b}^2
(1-r-\hat s) v^2 (\vel A_1 \ver^2 - \vel B_1 \ver^2 + \vel D_1
\ver^2 - \vel E_1 \ver^2 )
\Big\} \nnb \\
\ek \frac{1}{3} m_{\Lambda_b}^2 \{ 12 m_\ell^2 (1-r) +
m_{\Lambda_b}^2 \hat s [3 (1-r+\hat s) + v^2 (1-r-\hat s)] \}
(\vel A_2 \ver^2 - \vel B_2 \ver^2) \nnb \\
\ek \frac{2}{3} m_{\Lambda_b}^4 \hat s (2 - 2 r + \hat s) v^2 \,
(\vel D_2 \ver^2 - \vel E_2 \ver^2) \Bigg\}~,  \eea

\bea \label{e12} \lefteqn{ P_T(\hat s,
m_{t\prime},r_{sb},\phi_{sb}) = - \frac{8 \pi m_{\Lambda_b}^3 v
\sqrt{\hat s\lambda}} {\Delta(\hat
s,m_{t\prime},r_{sb},\phi_{sb})} \Bigg\{ m_\ell \Big( \mbox{\rm
Im}[(A_1+B_1)^\ast F_1]
\Big)} \nnb \\
\ek m_\ell m_{\Lambda_b} \Big[
(1+\sqrt{r}) \, \mbox{\rm Im}[(A_2+B_2)^\ast F_1] \Big] \nnb \\
\ar m_{\Lambda_b}^2 (1-r+\hat s) \Big( \mbox{\rm Im}[A_2^\ast D_1
- A_1^\ast D_2] -
\mbox{\rm Im}[B_2^\ast E_1 - B_1^\ast E_2] \Big)\nnb \\
\ar 2 m_{\Lambda_b} \Big( \mbox{\rm Im}[A_1^\ast E_1-B_1^\ast D_1]
- m_{\Lambda_b}^2 \hat s \, \mbox{\rm Im}[A_2^\ast E_2 - B_2^\ast
D_2] \Big)\Bigg\}~, \eea
The dependence of the $P_L$, $P_N$ and $P_T$
polarizations of the $\Lambda$ baryon on $t^\prime$ quark mass  at $\hat s=0.5$  and at three fixed values of the $r_{sb}$ and SM  are shown in figures 6-11.
\begin{figure} [h!]
\label{figPOLNm} \centering
\begin{tabular}{cc}
\epsfig{file=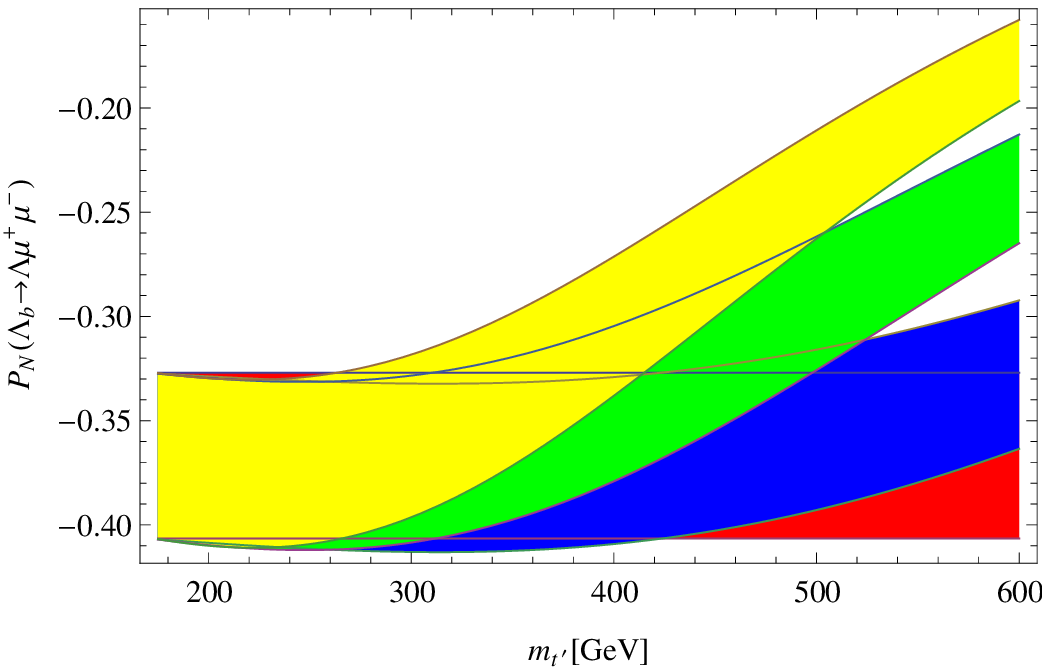,width=0.45\linewidth,clip=} &
\epsfig{file=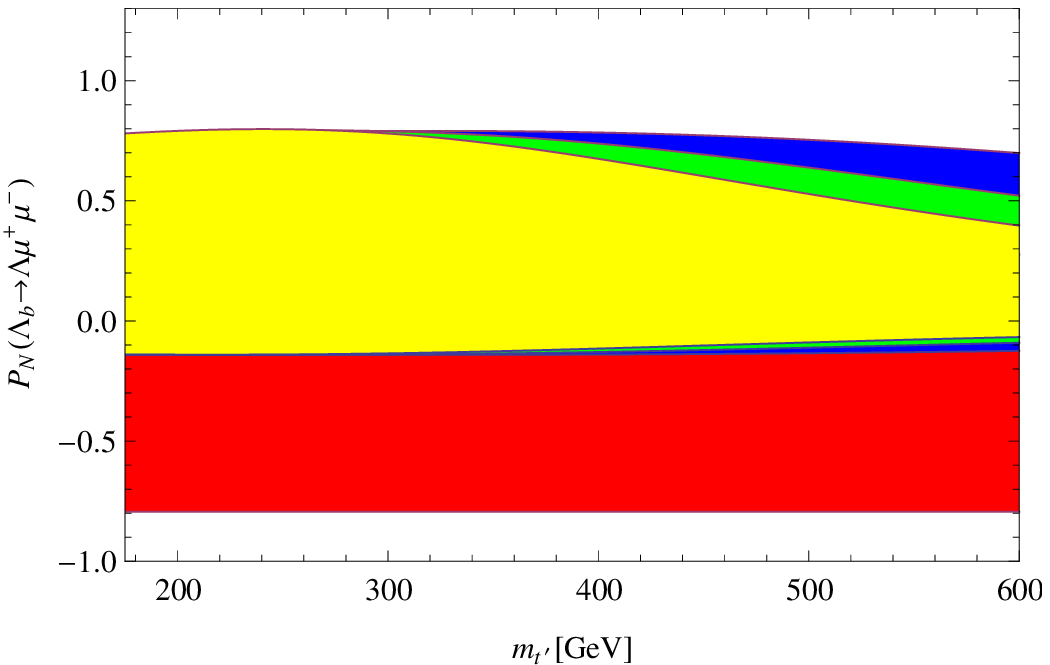,width=0.45\linewidth,clip=}
\end{tabular}
\caption{The dependence of normal baryon polarization  for the $\Lambda_b
\rar \Lambda \mu^+ \mu^-$ decay on $m_{t'}$ at $\hat s=0.5$. The red  band corresponds to  the  SM, while  the blue, green and yellow bands belong 
to the SM4 for  $r_{sb} = 0.005, ~0.01$ and $0.015$, respectively. The left graph corresponds to the HQET while the graph on the 
right refers to the full QCD.}
\end{figure}
\begin{figure} [h!]
 \label{figPOLN} \centering
\begin{tabular}{cc}
\epsfig{file=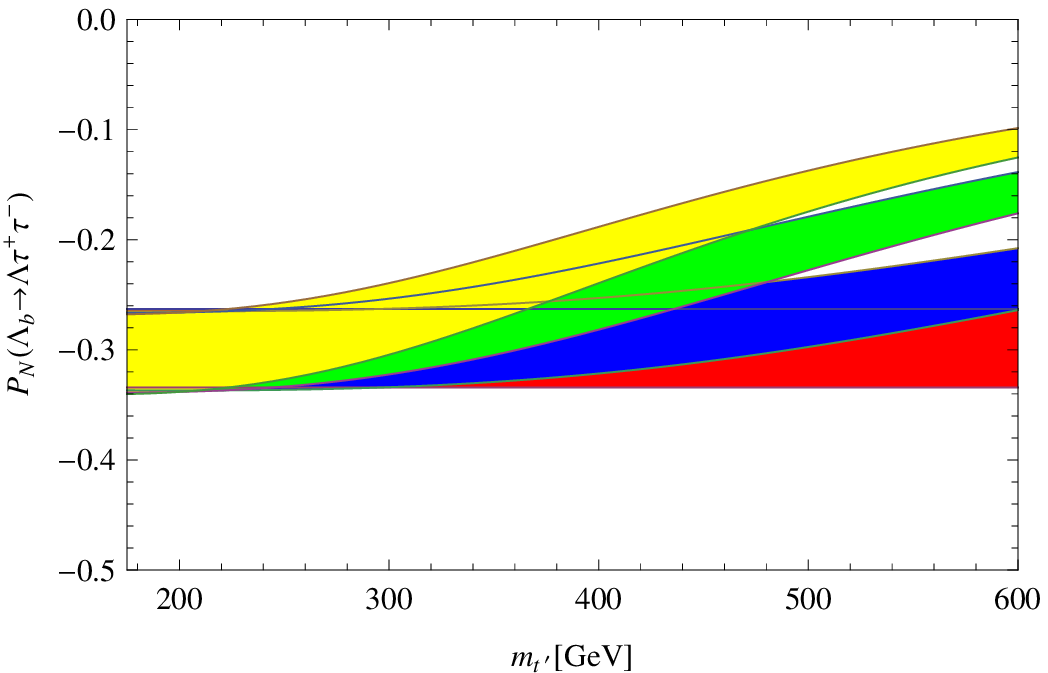,width=0.45\linewidth,clip=} &
\epsfig{file=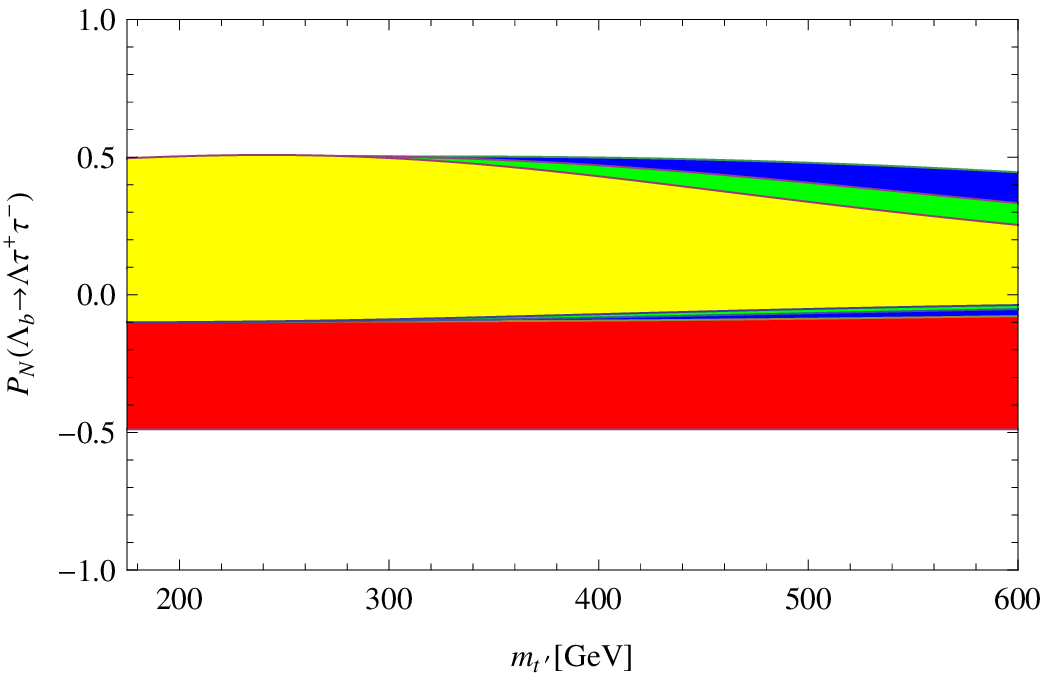,width=0.45\linewidth,clip=}
\end{tabular}
\caption{The same as FIG. 6 but for $\tau$. }
\end{figure}
\begin{figure}[h!]
\label{figPOLl} \centering
\begin{tabular}{cc}
\epsfig{file=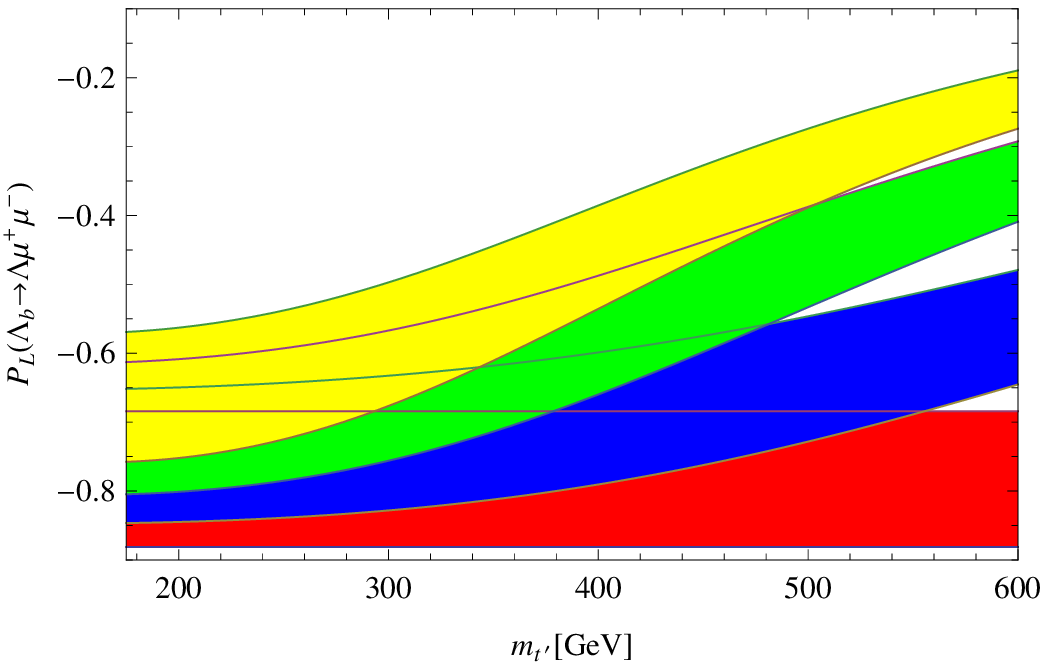,width=0.45\linewidth,clip=} &
\epsfig{file=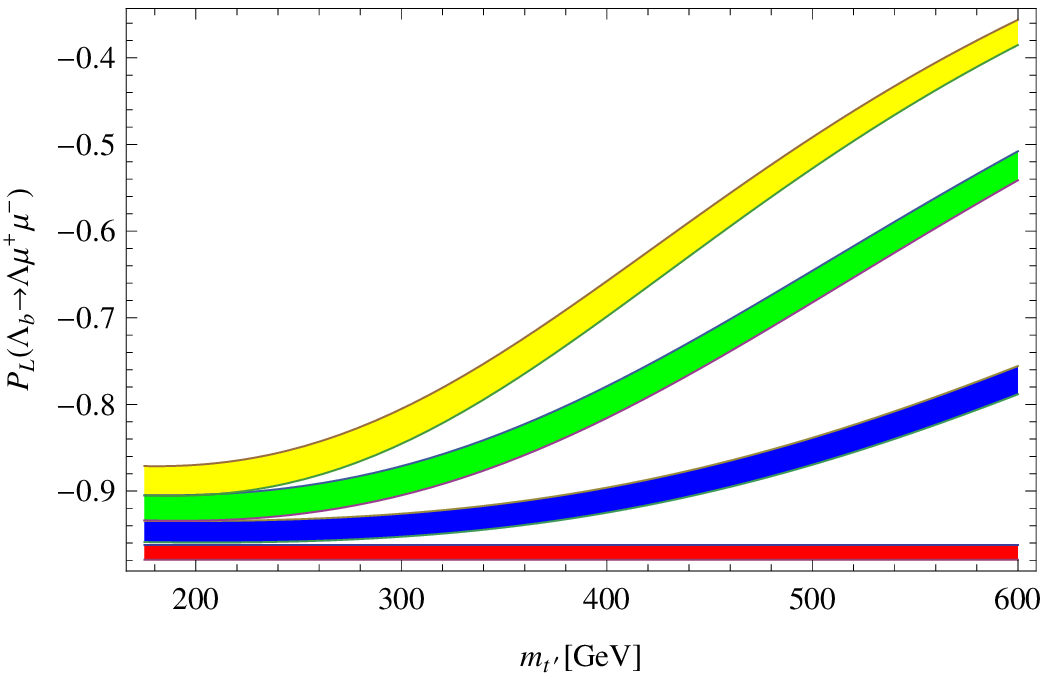,width=0.45\linewidth,clip=}
\end{tabular}
\caption{The same as FIG. 5 but for  longitudinal baryon polarization. }
\end{figure}
\begin{figure} [h!]
\label{figPOLl} \centering
\begin{tabular}{cc}
\epsfig{file=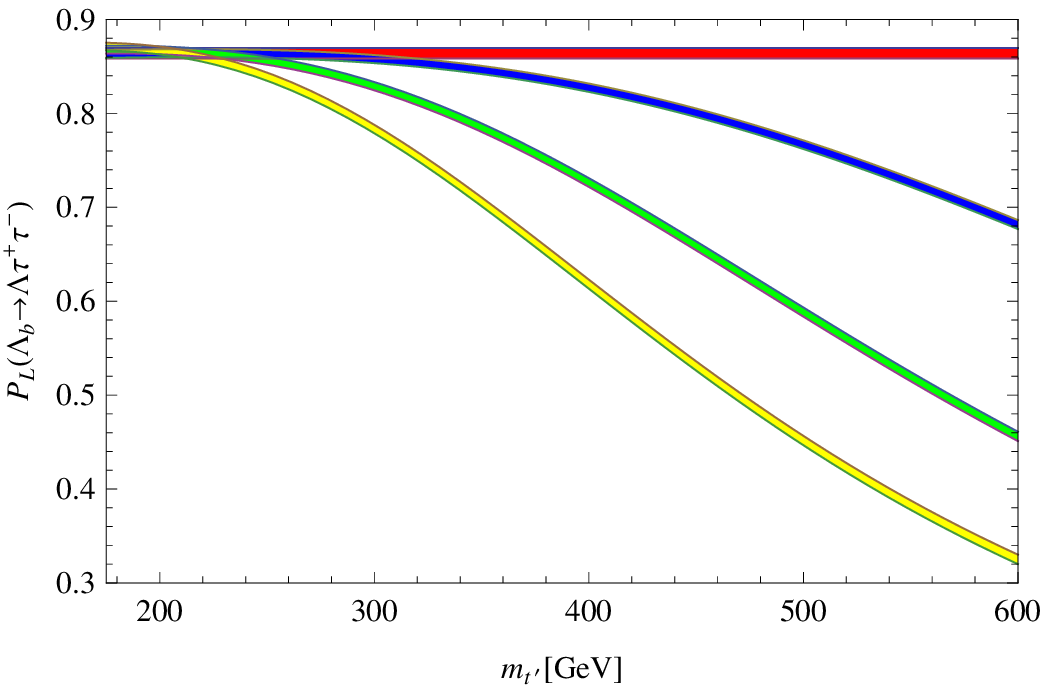,width=0.45\linewidth,clip=} &
\epsfig{file=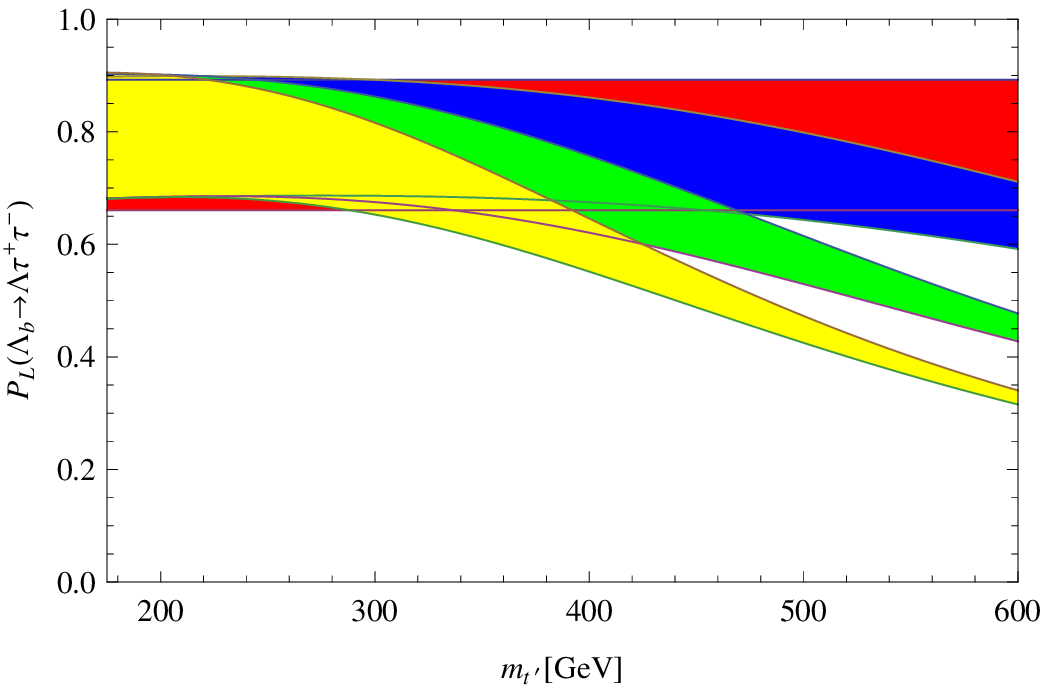,width=0.45\linewidth,clip=}
\end{tabular}
\caption{The same as FIG. 8 but for $\tau$. }
\end{figure}
\begin{figure} [h!]
\label{figPOLtm} \centering
\begin{tabular}{cc}
\epsfig{file=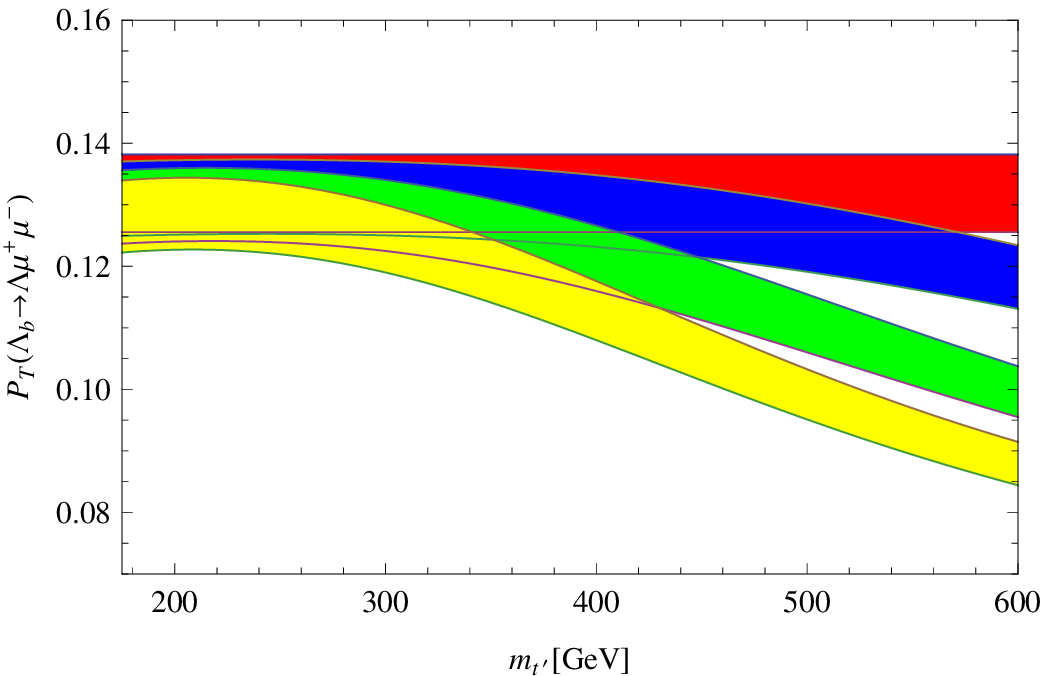,width=0.45\linewidth,clip=} &
\epsfig{file=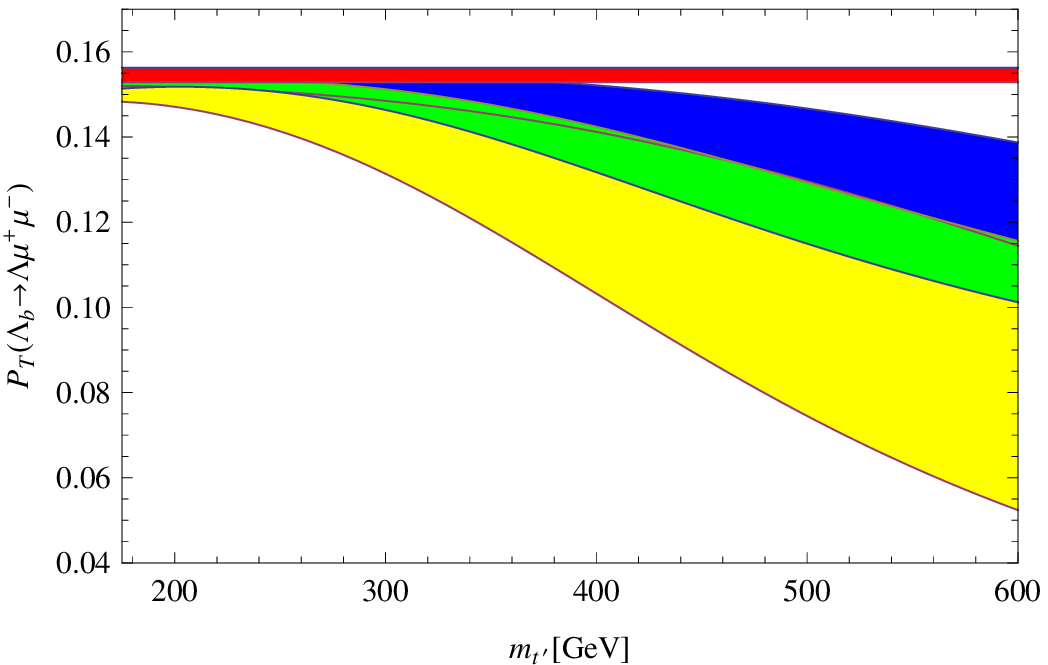,width=0.45\linewidth,clip=}
\end{tabular}
\caption{ The same as FIG. 5 but for  transverse baryon polarization. }
\end{figure}
\begin{figure}[h!]
\label{figPOLt} \centering
\begin{tabular}{cc}
\epsfig{file=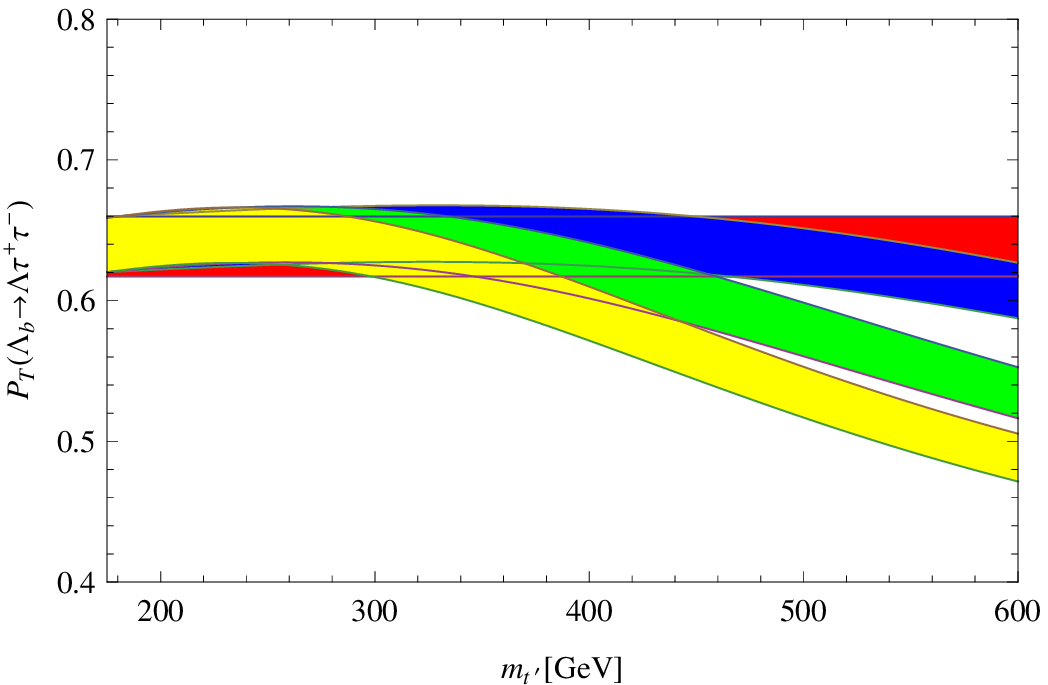,width=0.45\linewidth,clip=} &
\epsfig{file=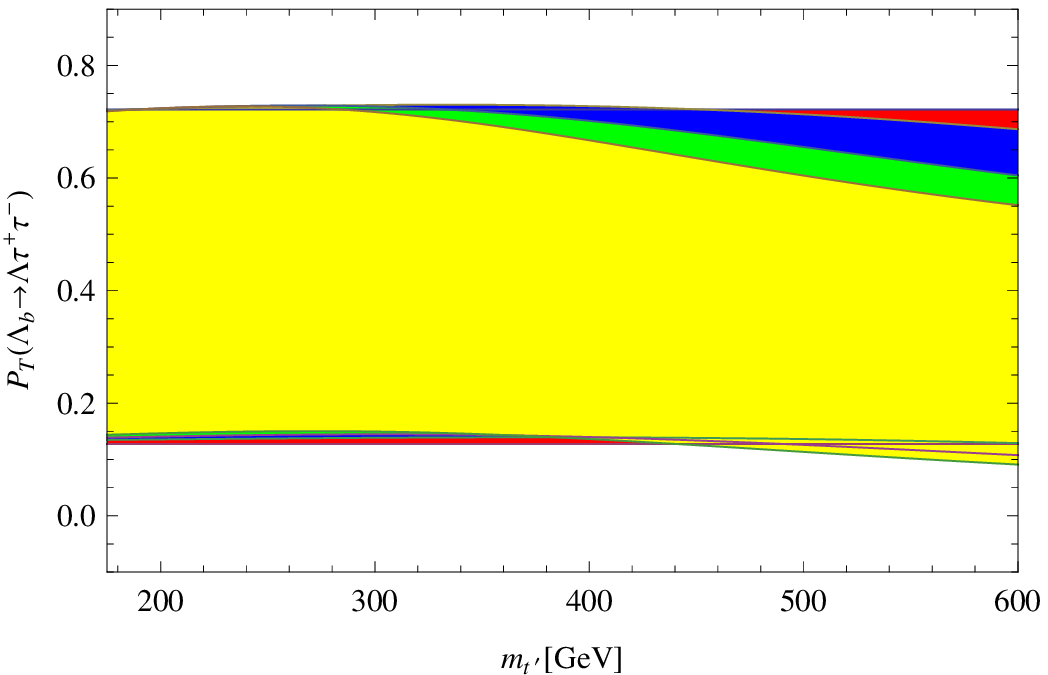,width=0.45\linewidth,clip=}
\end{tabular}
\caption{The same as FIG. 10 but for $\tau$.  }
\end{figure}

A quick glance in  the  figures 6-11 leads to the following conclusions:
\begin{itemize}
\item The baryon polarizations also overall  favor the  $m_{t'}\simeq 400~GeV$ for the lower limit of the fourth family quark.
\item Our numerical analysis show that as far as the central values of the form factors are considered, there are considerable differences between the full theory predictions on the  $P_N$ and  $P_T$ and the HQET results for both lepton channels. This difference is small for the $P_L$ and $\mu$ channel 
and is approximately  zero for the longitudinal polarization and $\tau$ channel. When we consider the uncertainties of the form factors we detect sizable differences in both values and behavior of the baryon polarizations
with respect to the $m_{t'}$ for all cases.
\item Except the full QCD predictions on the $P_N$ for both leptons and the $P_T$ for the $\tau$ channel,  the difference between the predictions of SM and SM4 grows with increasing the fourth generation 
quark mass.
 This difference also increases with increasing the value of the   $r_{sb}$. In the $P_N$ for both leptons and the $P_T$ for the $\tau$ channel, the uncertainties of the form factors lead to a very small difference 
between two model predictions.
\item When we consider only the central values of the form factors, the $|P_N|$ in $\mu$ channel is larger than that of $\tau$ at any values of the fourth generation parameters. The situation is inverse in the case of $|P_T|$. The $|P_L|$ is approximately the same for both lepton channels.
\end{itemize}

\subsection{Double Lepton Polarization Asymmetries} 
For the  general model independent form of the effective Hamiltonian, the double lepton
polarization asymmetries characterizing the considered decay channel 
are calculated in \cite{asikveli}. In the case of SM4, they reduce to the following explicit expressions in the rest frame of the $l^\pm$ (see also \cite{0608143,R7710}):

\bea \label{e7717} P_{LN}(\hat s,m_{t^\prime}, r_{sb},\phi_{sb})
 ~\es~ \frac{16 \pi m_{\Lambda_b}^4 \hat{m}_\ell
\sqrt{\lambda}}{\Delta(\hat s,m_{t^\prime}, r_{sb},\phi_{sb})
\sqrt{\hat{s}}} \mbox{\rm Im} \Bigg\{
%1
(1-r) (A_1^\ast D_1 + B_1^\ast E_1)
%2
%3
%4
+ m_{\Lambda_b}
 \hat{s} (A_1^\ast E_3 - A_2^\ast E_1 \nnb \\
 + B_1^\ast D_3
-B_2^\ast D_1)
%5
%6
%7
%8
%9
\ar m_{\Lambda_b}
 \sqrt{r} \hat{s}
(A_1^\ast D_3 + A_2^\ast D_1+B_1^\ast E_3 + B_2^\ast E_1)
%10
%11
- m_{\Lambda_b}^2 \hat{s}^2 \Big( B_2^\ast E_3 + A_2^\ast D_3
\Big) \Bigg\},\eea

\bea \label{e7719} P_{LT}(\hat s,m_{t^\prime}, r_{sb},\phi_{sb})
~\es~ \frac{16 \pi m_{\Lambda_b}^4 \hat{m}_\ell \sqrt{\lambda}
v}{\Delta(\hat s,m_{t^\prime}, r_{sb},\phi_{sb}) \sqrt{\hat{s}}}
\mbox{\rm Re} \Bigg\{
%1
(1-r) \Big( \vel D_1 \ver^2 + \vel E_1 \ver^2 \Big)
%2
- \hat{s} \Big(A_1 D_1^\ast - B_1 E_1^\ast \Big) \nnb \\
%3
\ek m_{\Lambda_b} \hat{s} \Big[ B_1 D_2^\ast + (A_2 + D_2 -D_3)
E_1^\ast -  A_1 E_2^\ast
-(B_2-E_2+E_3) D_1^\ast \Big] \nnb \\
%4
%5
%6
%8
\ar m_{\Lambda_b}
 \sqrt{r} \hat{s}
\Big[ A_1 D_2^\ast + (A_2 + D_2 +D_3) D_1^\ast - B_1 E_2^\ast -
(B_2 - E_2 - E_3) E_1^\ast \Big] \nnb \\
%7
\ar m_{\Lambda_b}^2 \hat{s} (1-r) (A_2 D_2^\ast - B_2 E_2^\ast)
%9
- m_{\Lambda_b}^2 \hat{s}^2 (D_2 D_3^\ast + E_2 E_3^\ast )\Bigg\},
\eea

\bea \label{e7721} P_{NT} (\hat s,m_{t^\prime},
r_{sb},\phi_{sb})~\es~ \frac{64 m_{\Lambda_b}^4 \lambda v}{3
\Delta(\hat s,m_{t^\prime}, r_{sb},\phi_{sb})} \mbox{\rm Im}
\Bigg\{
%1
(A_1 D_1^\ast +B_1 E_1^\ast)
%2
%3
%4
%5
%6
%7
+ m_{\Lambda_b}^2 \hat{s} (A_2^\ast D_2 + B_2^\ast E_2)\Bigg\},
%8
%9
\eea

\bea \label{e7723} P_{NN} (\hat s,m_{t^\prime},
r_{sb},\phi_{sb})~\es~ \frac{32 m_{\Lambda_b}^4}{3 \hat{s}
\Delta(\hat s,m_{t^\prime}, r_{sb},\phi_{sb})} \mbox{\rm Re}
\Bigg\{
%1
%2
24 \hat{m}_\ell^2 \sqrt{r} \hat{s}
( A_1 B_1^\ast + D_1 E_1^\ast ) \nnb \\
%3
\ek 12 m_{\Lambda_b} \hat{m}_\ell^2 \sqrt{r} \hat{s}
(1-r +\hat{s}) (A_1 A_2^\ast + B_1 B_2^\ast) \nnb \\
%4
%5
%6
%7
%8
%9
\ar 6 m_{\Lambda_b} \hat{m}_\ell^2 \hat{s} \Big[ m_{\Lambda_b}
\hat{s} (1+r-\hat{s}) \Big(\vel D_3 \ver^2 + \vel E_3 \ver^2 \Big)
+ 2 \sqrt{r} (1-r+\hat{s})
(D_1 D_3^\ast + E_1 E_3^\ast)\Big] \nnb \\
%10
\ar 12 m_{\Lambda_b} \hat{m}_\ell^2 \hat{s} (1-r-\hat{s})
(A_1 B_2^\ast + A_2 B_1 ^\ast + D_1 E_3^\ast + D_3 E_1^\ast) \nnb \\
%11
\ek [ \lambda \hat{s} + 2 \hat{m}_\ell^2 (1 + r^2 - 2 r + r
\hat{s} + \hat{s} - 2 \hat{s}^2) ] \Big( \vel A_1 \ver^2 + \vel
B_1 \ver^2 - \vel D_1 \ver^2 -
\vel E_1 \ver^2 \Big) \nnb \\
%12
%13
\ar 24 m_{\Lambda_b}^2 \hat{m}_\ell^2 \sqrt{r} \hat{s}^2 (A_2
B_2^\ast + D_3 E_3^\ast)
%14
%15
- m_{\Lambda_b}^2 \lambda \hat{s}^2 v^2
\Big( \vel D_2 \ver^2 + \vel E_2 \ver^2 \Big) \nnb \\
%16
\ar m_{\Lambda_b}^2 \hat{s} \{ \lambda \hat{s} - 2 \hat{m}_\ell^2
[2 (1+ r^2) - \hat{s} (1+\hat{s}) - r (4+\hat{s})]\} \Big( \vel
A_2 \ver^2 + \vel B_2 \ver^2 \Big)\Bigg\},
%17
%18
\eea

\bea \label{e7724} P_{TT}(\hat s,m_{t^\prime}, r_{sb},\phi_{sb})
~\es~ \frac{32 m_{\Lambda_b}^4}{3 \hat{s} \Delta(\hat
s,m_{t^\prime}, r_{sb},\phi_{sb})} \mbox{\rm Re} \Bigg\{
%1
%2
- 24 \hat{m}_\ell^2 \sqrt{r} \hat{s}
( A_1 B_1^\ast + D_1 E_1^\ast ) \nnb \\
%3
\ek 12 m_{\Lambda_b} \hat{m}_\ell^2 \sqrt{r} \hat{s} (1-r
+\hat{s}) (D_1 D_3^\ast + E_1 E_3^\ast)
%4
%5
%6
- 24 m_{\Lambda_b}^2 \hat{m}_\ell^2 \sqrt{r} \hat{s}^2
( A_2 B_2^\ast + D_3 E_3^\ast ) \nnb \\
%7
%8
%9
\ek 6 m_{\Lambda_b} \hat{m}_\ell^2 \hat{s} \Big[ m_{\Lambda_b}
\hat{s} (1+r-\hat{s}) \Big(\vel D_3 \ver^2 + \vel E_3 \ver^2 \Big)
- 2 \sqrt{r} (1-r+\hat{s})
(A_1 A_2^\ast + B_1 B_2^\ast)\Big] \nnb \\
%10
\ek 12 m_{\Lambda_b} \hat{m}_\ell^2 \hat{s} (1-r-\hat{s})
(A_1 B_2^\ast + A_2 B_1 ^\ast + D_1 E_3^\ast + D_3 E_1^\ast) \nnb \\
%11
\ek [ \lambda \hat{s} - 2 \hat{m}_\ell^2 (1 + r^2 - 2 r + r
\hat{s} + \hat{s} - 2 \hat{s}^2) ]
\Big( \vel A_1 \ver^2 + \vel B_1 \ver^2 \Big) \nnb \\
%12
\ar m_{\Lambda_b}^2 \hat{s} \{ \lambda \hat{s} + \hat{m}_\ell^2 [4
(1- r)^2 - 2 \hat{s} (1+r) - 2 \hat{s}^2 ]\}
\Big( \vel A_2 \ver^2 + \vel B_2 \ver^2 \Big) \nnb \\
%13
\ar \{ \lambda \hat{s} - 2 \hat{m}_\ell^2 [5 (1- r)^2 - 7 \hat{s}
(1+r) + 2 \hat{s}^2 ]\}
\Big( \vel D_1 \ver^2 + \vel E_1 \ver^2 \Big) \nnb \\
\ek m_{\Lambda_b}^2 \lambda \hat{s}^2 v^2 \Big( \vel D_2 \ver^2 +
\vel E_2 \ver^2 \Big) \Bigg\}, \eea

 \bea \label{e7716} P_{LL}(\hat s,m_{t^\prime}, r_{sb},\phi_{sb}) ~\es~
\frac{16 m_{\Lambda_b}^4}{3\Delta(\hat s,m_{t^\prime},
r_{sb},\phi_{sb})}
\mbox{\rm Re} \Bigg\{\nnb \\
%1
%2
%3
%4
%5
%6
\ek 6 m_{\Lambda_b} \sqrt{r} (1-r+\hat{s}) \Big[ \hat{s} (1+v^2)
(A_1 A_2^\ast + B_1 B_2^\ast)  -
4 \hat{m}_\ell^2 (D_1 D_3^\ast + E_1 E_3^\ast) \Big] \nnb \\
%7
%8
\ar 6 m_{\Lambda_b} (1-r-\hat{s}) \Big[ \hat{s} (1+v^2) (A_1
B_2^\ast + A_2 B_1^\ast) +
4 \hat{m}_\ell^2 (D_1 E_3^\ast + D_3 E_1^\ast) \Big] \nnb \\
%9
\ar 12 \sqrt{r} \hat{s} (1+v^2) \Big( A_1 B_1^\ast + D_1 E_1^\ast
+
m_{\Lambda_b}^2 \hat{s} A_2 B_2^\ast \Big) \nnb \\
%10
\ar 12 m_{\Lambda_b}^2 \hat{m}_\ell^2 \hat{s} (1+r-\hat{s})
\ga \vel D_3 \ver^2 + \vel E_3^\ast \ver^2 \dr \nnb \\
%11
%12
\ek (1+v^2) \Big[ 1+r^2 - r (2-\hat{s}) +\hat{s} (1-2 \hat{s})
\Big]
\Big(\vel A_1 \ver^2 + \vel B_1 \ver^2 \Big) \nnb \\
%13
\ek \Big[ (5 v^2 - 3) (1-r)^2 + 4 \hat{m}_\ell^2 (1+r) + 2 \hat{s}
(1+8 \hat{m}_\ell^2 + r)
- 4 \hat{s}^2 \Big] \Big( \vel D_1 \ver^2 + \vel E_1 \ver^2 \Big) \nnb \\
%14
\ek m_{\Lambda_b}^2 (1+v^2) \hat{s} \Big[2 + 2 r^2 -\hat{s}(1
+\hat{s}) - r (4 + \hat{s})\Big] \big(
\vel A_2 \ver^2 + \vel B_2 \ver^2 \Big) \nnb \\
%15
%16
\ek 2 m_{\Lambda_b}^2 \hat{s} v^2 \Big[ 2 (1 + r^2) - \hat{s}
(1+\hat{s}) - r (4+\hat{s})\Big] \Big(
\vel D_2 \ver^2 + \vel E_2 \ver^2 \Big) \nnb \\
%17
\ar 12 m_{\Lambda_b} \hat{s} (1-r-\hat{s}) v^2
\Big( D_1 E_2^\ast + D_2 E_1^\ast \Big) \nnb \\
%18
\ek 12 m_{\Lambda_b} \sqrt{r} \hat{s} (1-r+\hat{s}) v^2
\Big( D_1 D_2^\ast + E_1 E_2^\ast \Big) \nnb \\
%19
\ar 24 m_{\Lambda_b}^2 \sqrt{r} \hat{s} \Big( \hat{s} v^2 D_2
E_2^\ast + 2 \hat{m}_\ell^2 D_3 E_3^\ast \Big)\Bigg\}, \eea
 where $\hat
m_l=\frac{m_l}{m_{\Lambda_b}}$. 
Some of the double lepton polarization asymmetries as a function of the $m_{t^\prime}$   at $\hat s=0.5$  and at three fixed values of the  $r_{sb}$ and the SM  are shown in figures 12-21.
\begin{figure} [h!]
\label{POLlnm} \centering
\begin{tabular}{cc}
\epsfig{file=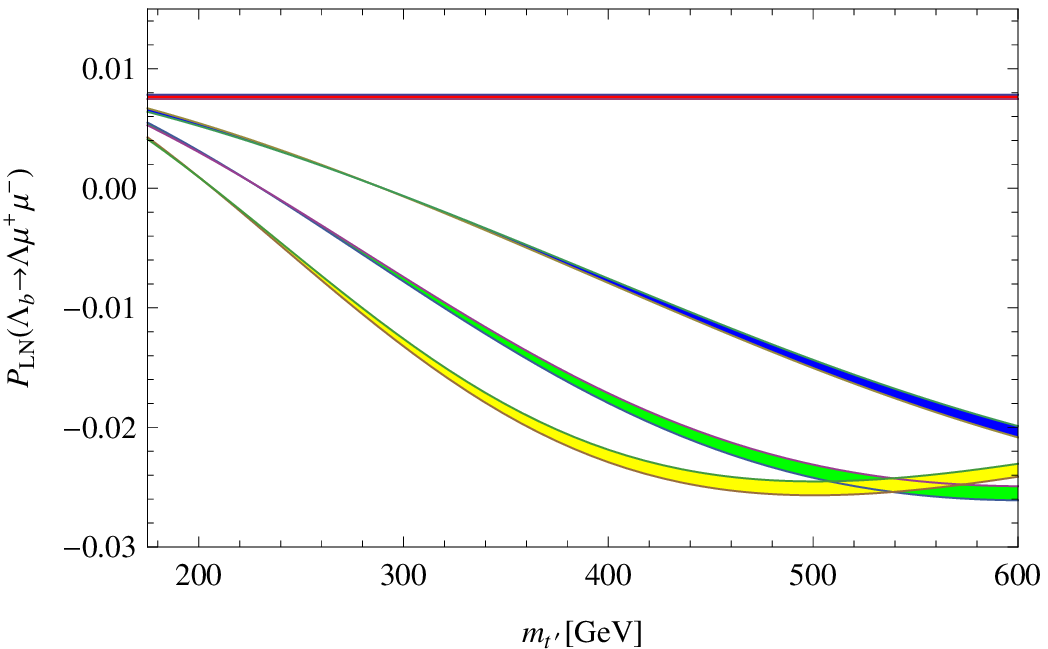,width=0.45\linewidth,clip=} &
\epsfig{file=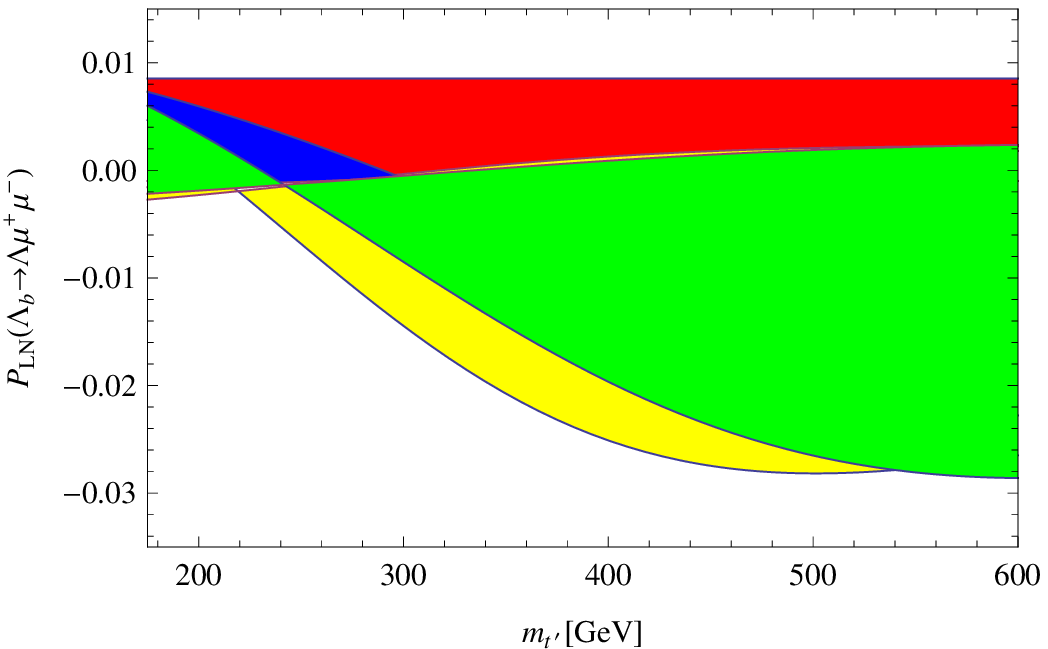,width=0.45\linewidth,clip=}
\end{tabular}
\caption{ The dependence of double lepton polarization asymmetry $P_{LN}$  for the $\Lambda_b \rar \Lambda \mu^+ \mu^-$ decay
on $m_{t'}$  at $\hat s=0.5$. The red  band corresponds to  the  SM, while  the blue, green and yellow bands belong 
to the SM4 for  $r_{sb} = 0.005, ~0.01$ and $0.015$, respectively. The left graph corresponds to the HQET while the graph on the 
right refers to the full QCD. }
\end{figure}
\begin{figure} [h!]
\label{POLln} \centering
\begin{tabular}{cc}
\epsfig{file=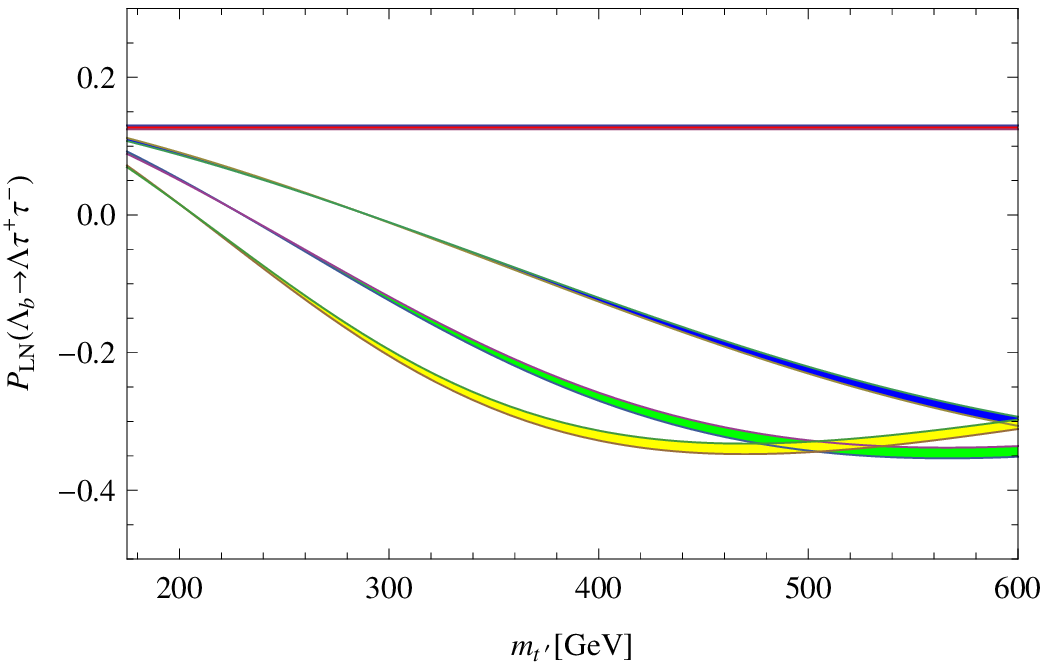,width=0.45\linewidth,clip=} &
\epsfig{file=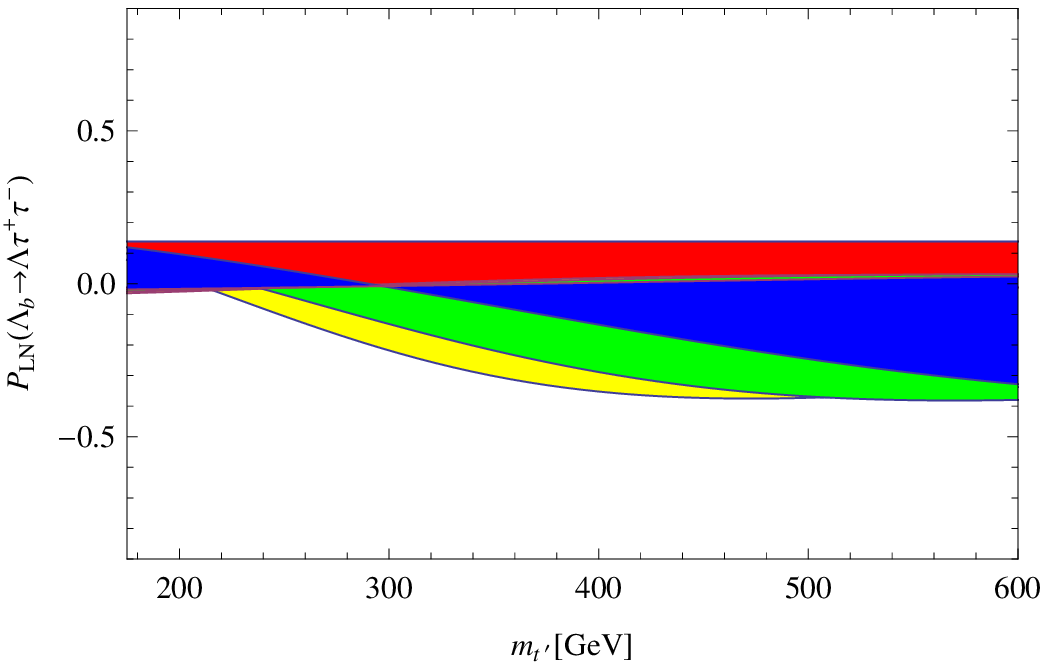,width=0.45\linewidth,clip=}
\end{tabular}
\caption{The same as FIG. 12 but for $\tau$. }
\end{figure}
\begin{figure} [h!]
\label{figltm} \centering
\begin{tabular}{cc}
\epsfig{file=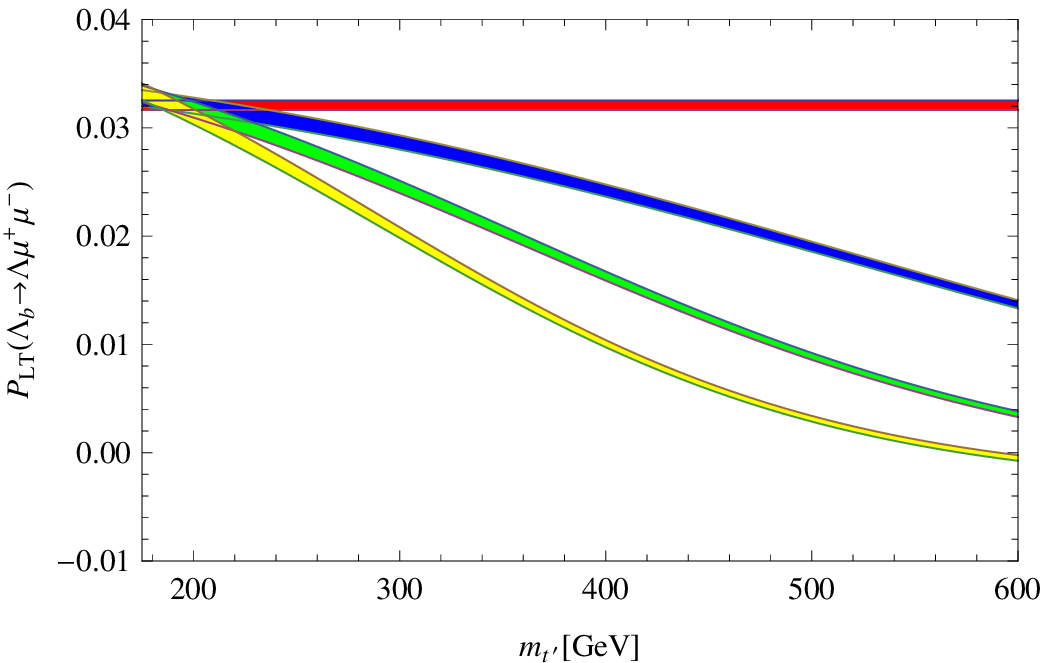,width=0.45\linewidth,clip=} &
\epsfig{file=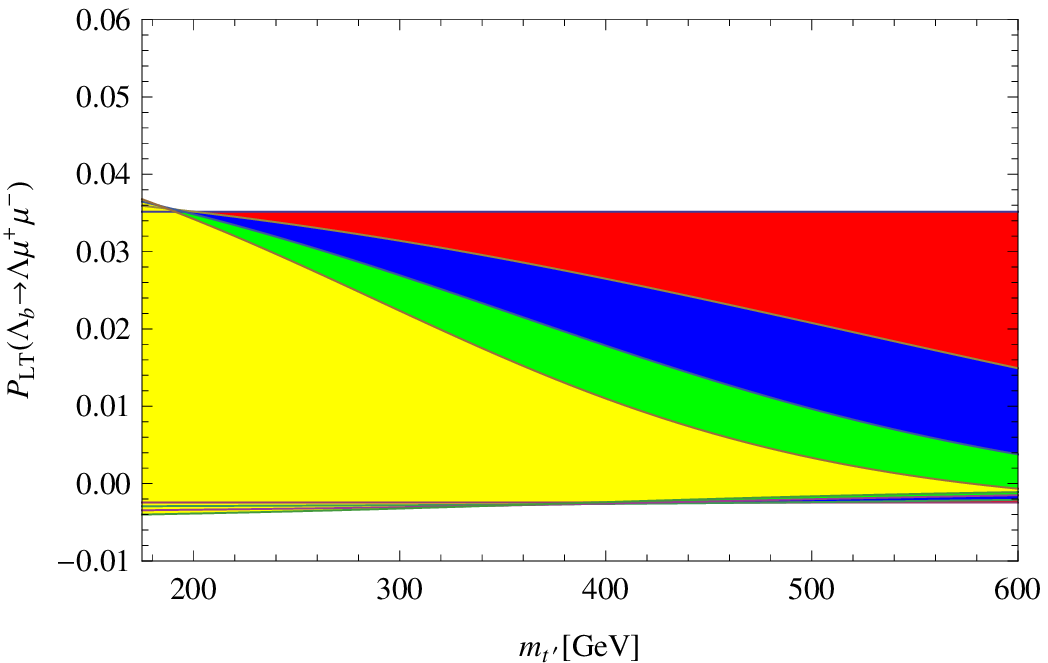,width=0.45\linewidth,clip=}
\end{tabular}
\caption{The same as FIG. 12 but for   $P_{LT}$. }
\end{figure}
\begin{figure} [h!]
\label{figlt} \centering
\begin{tabular}{cc}
\epsfig{file=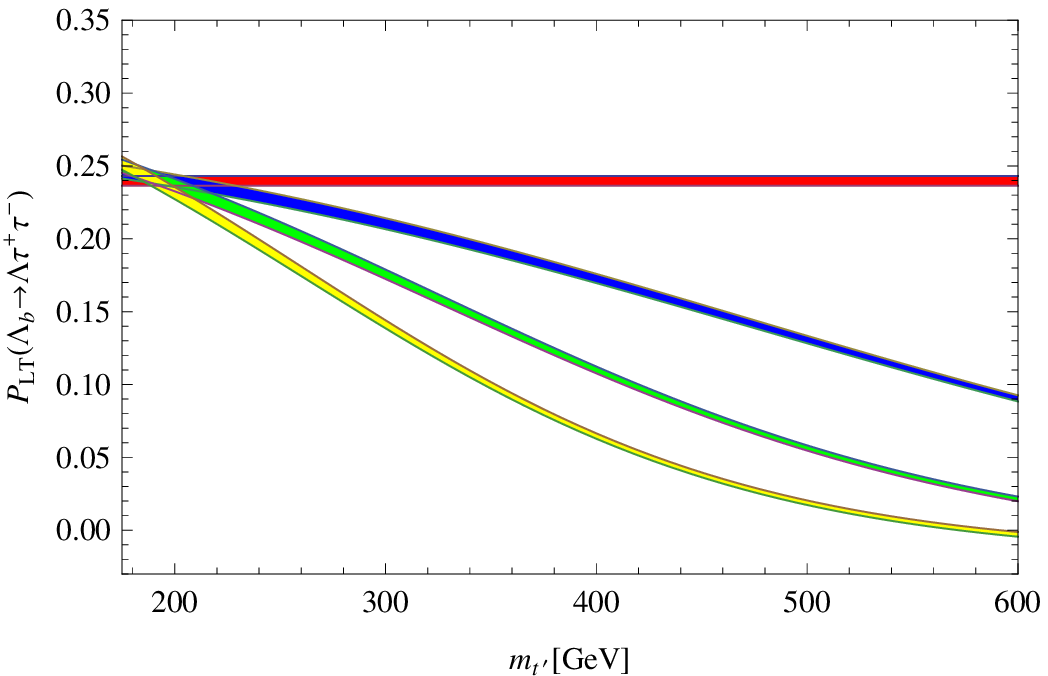,width=0.45\linewidth,clip=} &
\epsfig{file=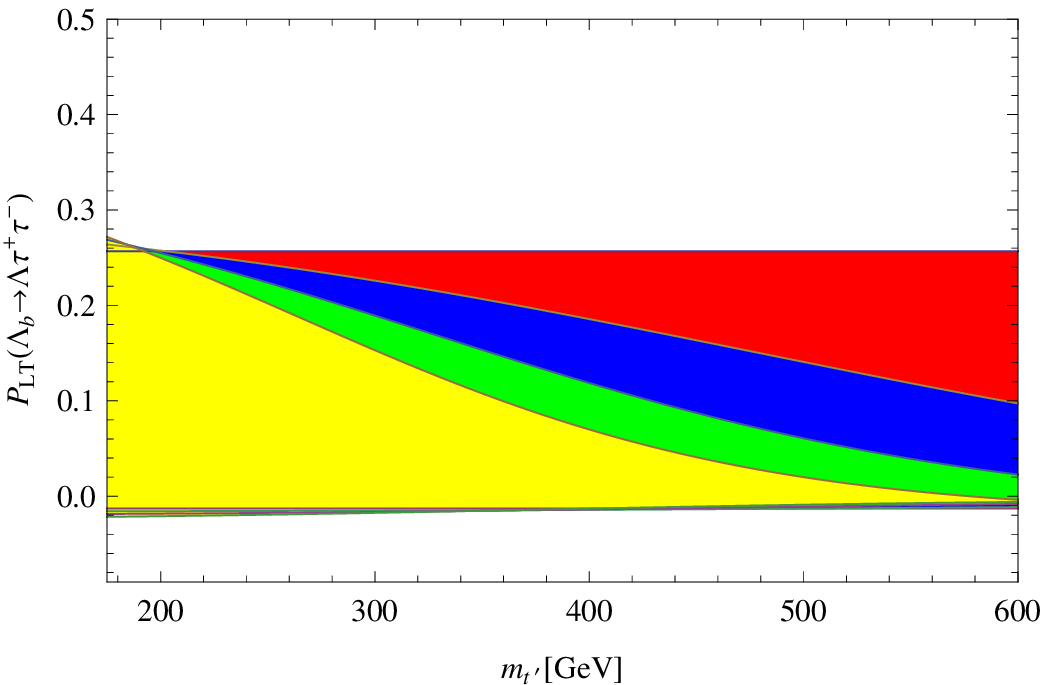,width=0.45\linewidth,clip=}
\end{tabular}
\caption{The same as FIG. 14 but for $\tau$.  }
\end{figure}
\begin{figure} [h!]
\label{fignTm} \centering
\begin{tabular}{cc}
\epsfig{file=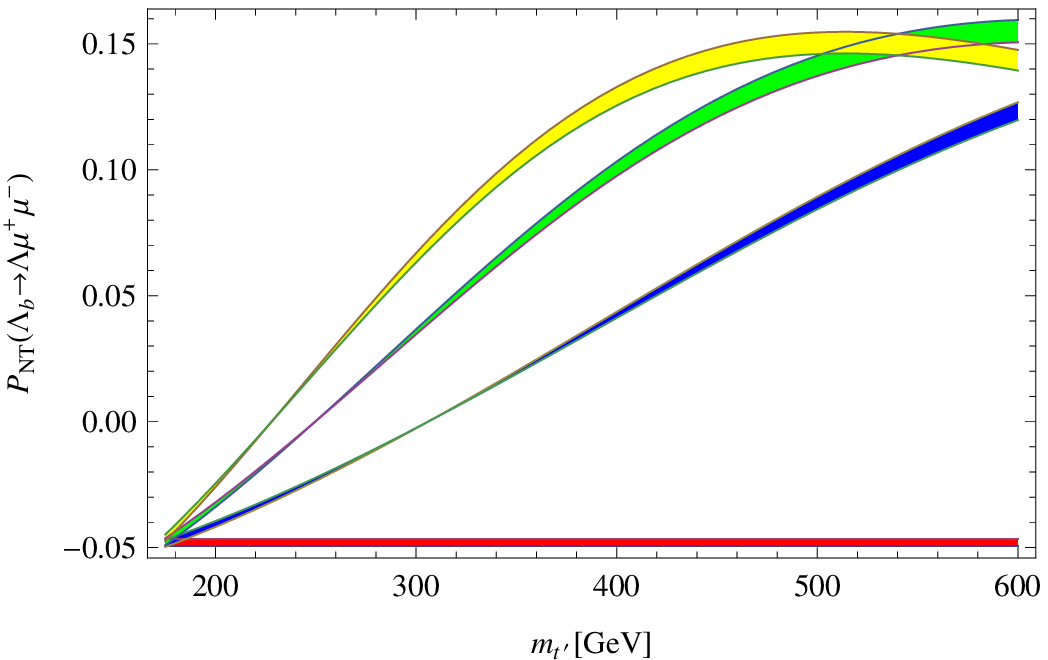,width=0.45\linewidth,clip=} &
\epsfig{file=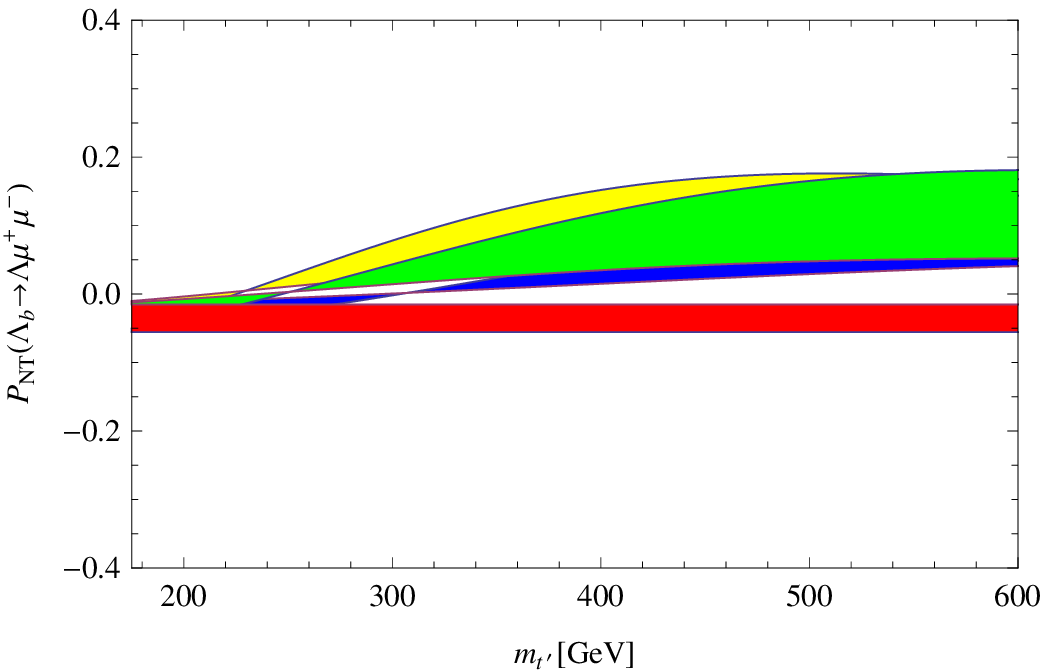,width=0.45\linewidth,clip=}
\end{tabular}
\caption{The same as FIG. 12 but for  $P_{NT}$. }
\end{figure}
\begin{figure} [h!]
\label{nT} \centering
\begin{tabular}{cc}
\epsfig{file=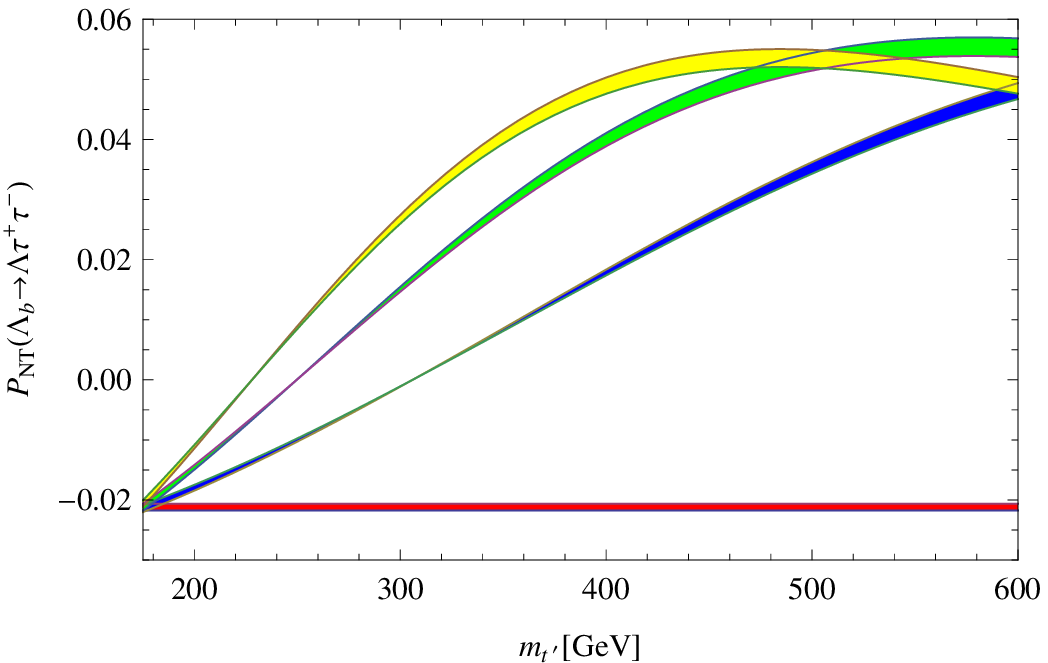,width=0.45\linewidth,clip=} &
\epsfig{file=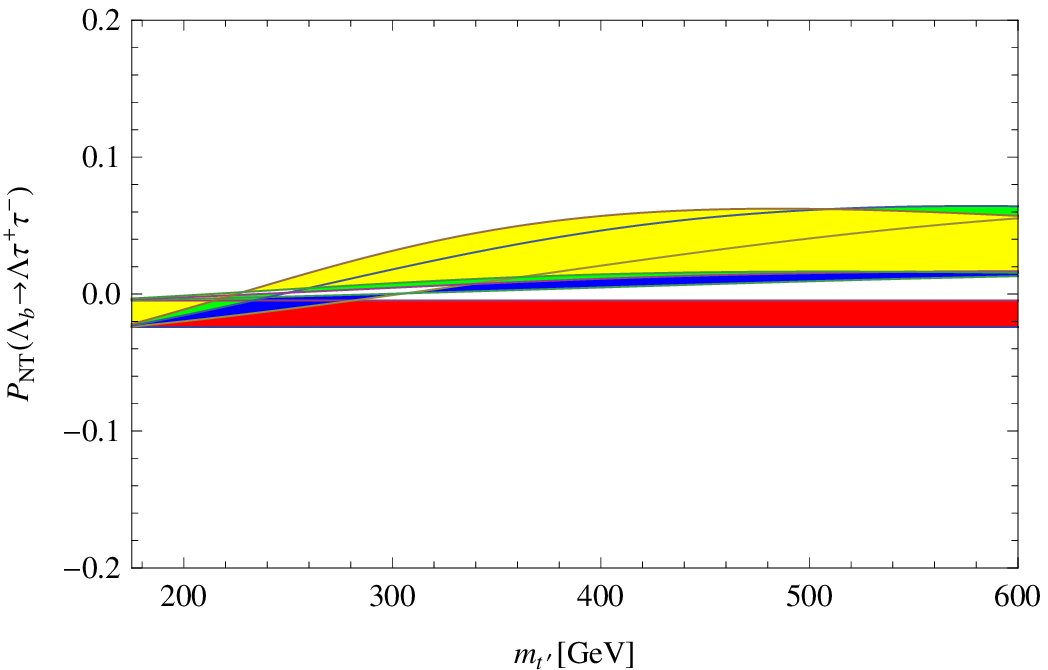,width=0.45\linewidth,clip=}
\end{tabular}
\caption{The same as FIG. 16 but for $\tau$. }
\end{figure}
\begin{figure} [h!]
\label{figPOLnnm} \centering
\begin{tabular}{cc}
\epsfig{file=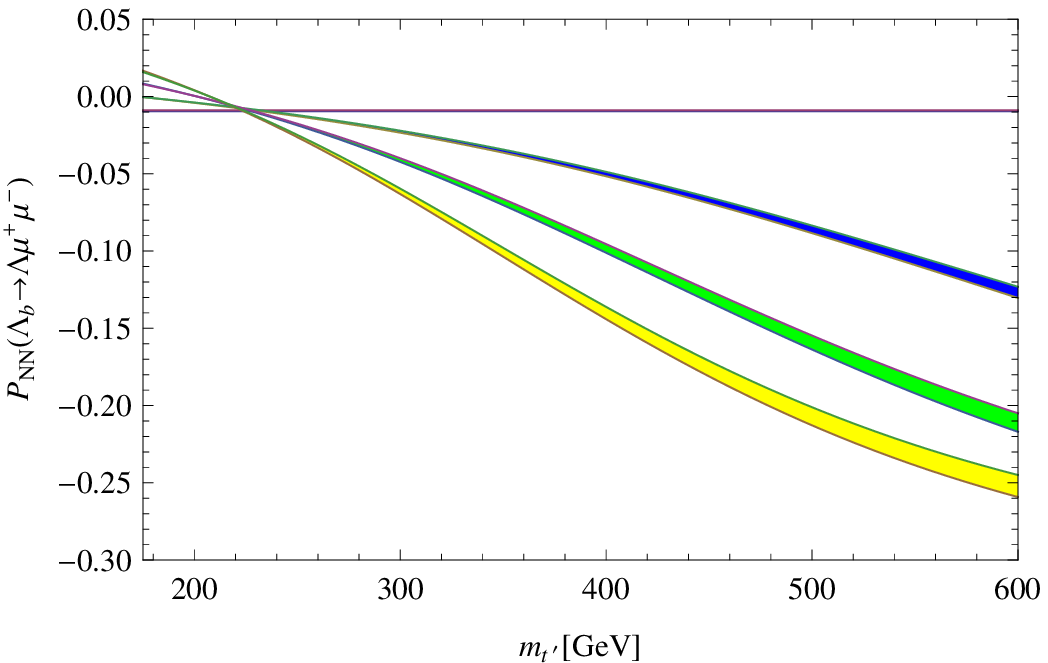,width=0.45\linewidth,clip=} &
\epsfig{file=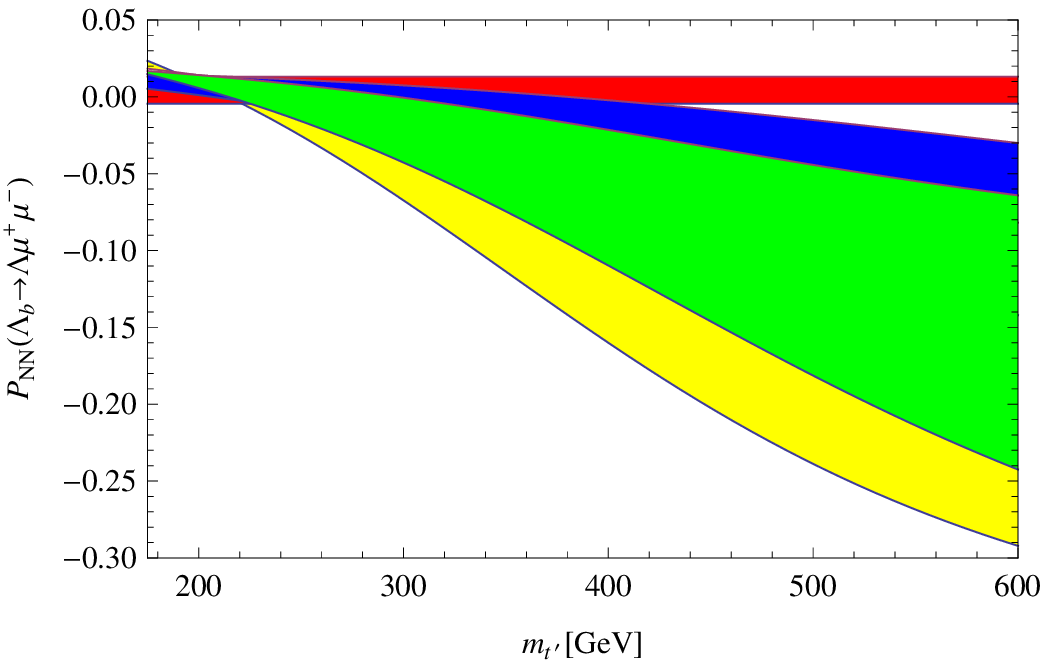,width=0.45\linewidth,clip=}
\end{tabular}
\caption{The same as FIG. 12 but for  $P_{NN}$. }
\end{figure}
\begin{figure} [h!]
 \label{fignn} \centering
\begin{tabular}{cc}
\epsfig{file=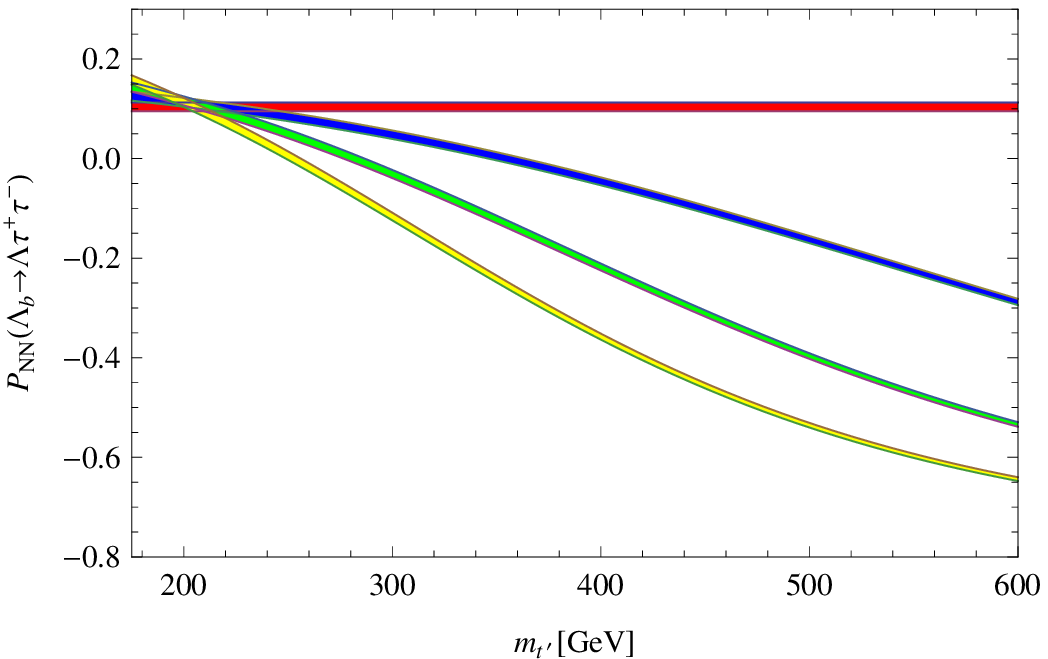,width=0.45\linewidth,clip=} &
\epsfig{file=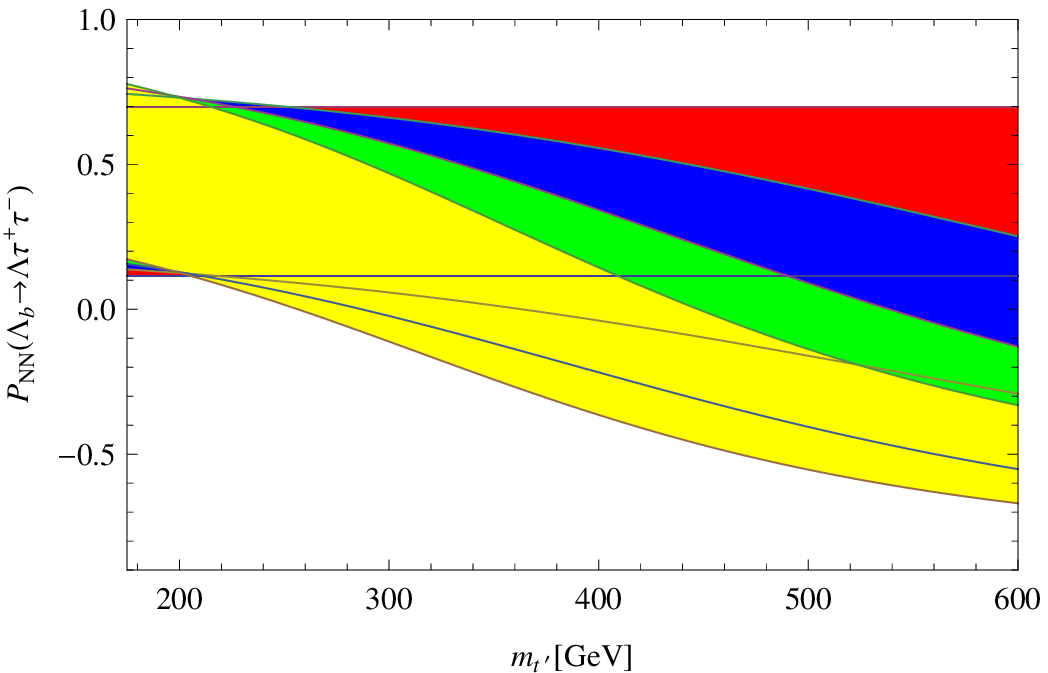,width=0.45\linewidth,clip=}
\end{tabular}
\caption{The same as FIG. 18 but for $\tau$.}
\end{figure}
\begin{figure} [h!]
\label{POLttm} \centering
\begin{tabular}{cc}
\epsfig{file=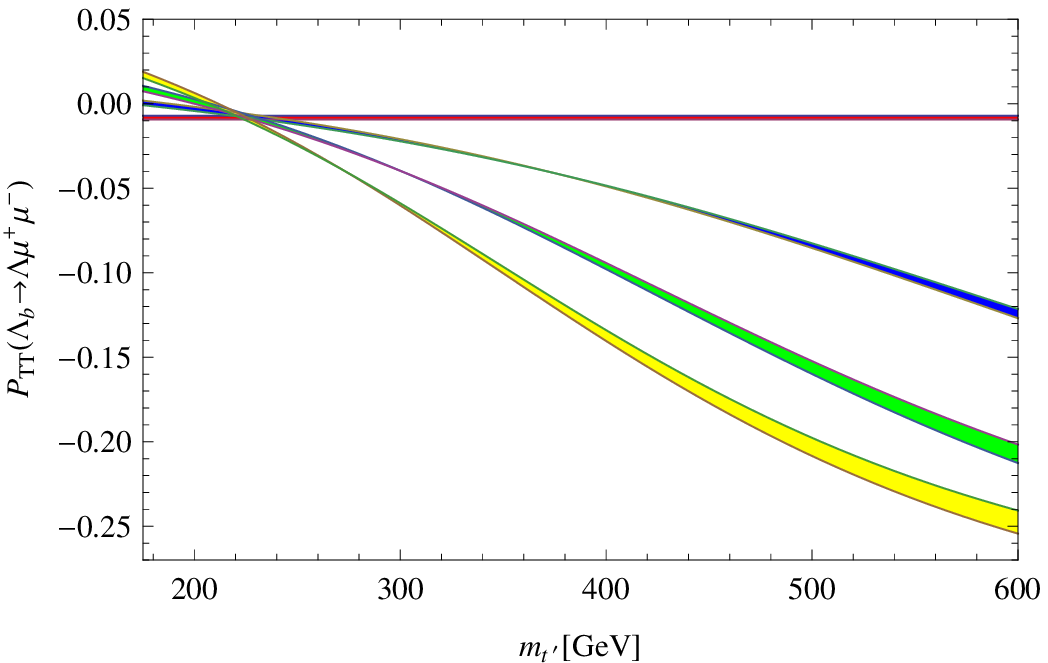,width=0.45\linewidth,clip=} &
\epsfig{file=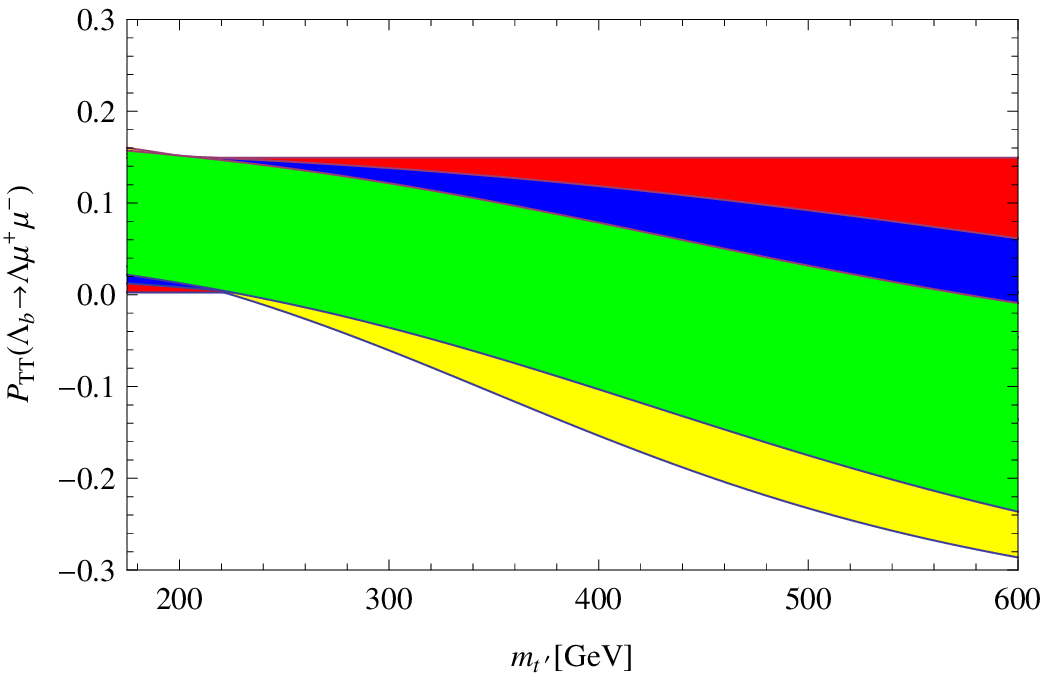,width=0.45\linewidth,clip=}
\end{tabular}
\caption{The same as FIG. 12 but for $P_{TT}$. }
\end{figure}
\begin{figure} [h!]
\label{tt} \centering
\begin{tabular}{cc}
\epsfig{file=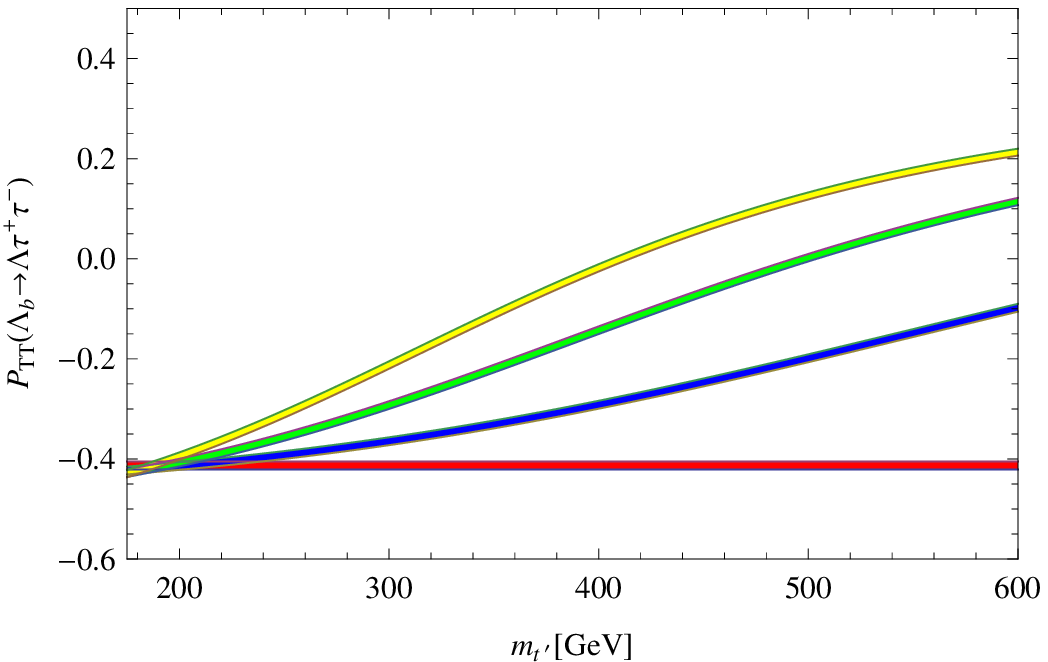,width=0.45\linewidth,clip=} &
\epsfig{file=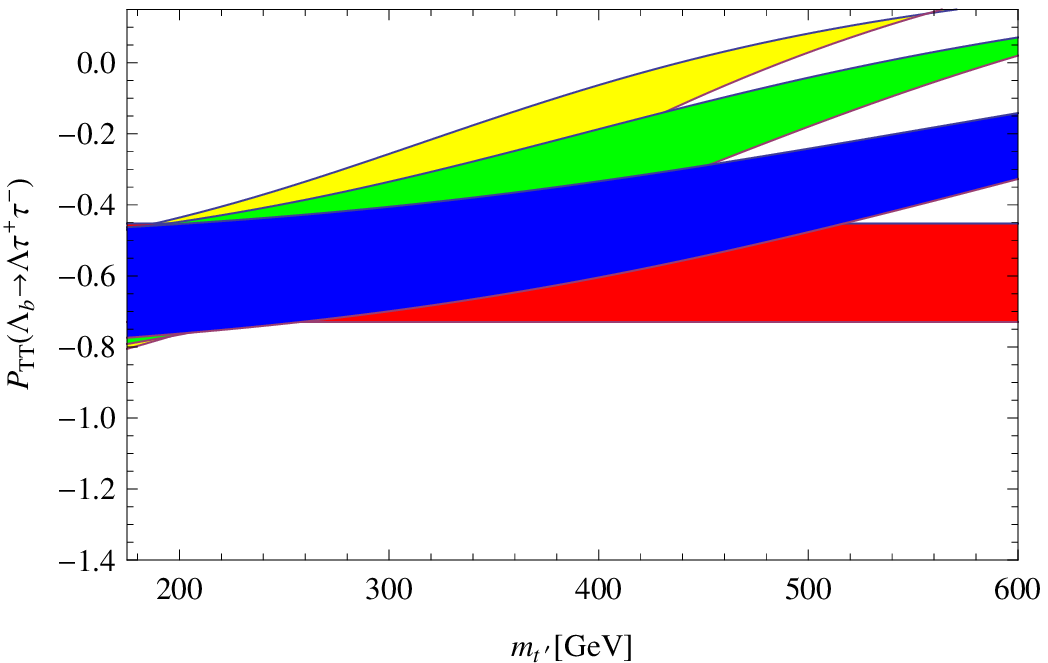,width=0.45\linewidth,clip=}
\end{tabular}
\caption{The same as FIG. 20 but for $\tau$.}
\end{figure}

From the analysis of the figures 12-21, we conclude the following items:
\begin{itemize}
\item When we consider only the central values of the form factors, our numerical results show that  there are  sizable differences between the full QCD and HQET results (HQET violation) 
in the $P_{TT}$ and $P_{NN}$ polarizations for $\tau$ channel and at fixed values of the fourth generation parameters.
The results of two models on $P_{LT}$, $P_{LN}$ and $P_{NT}$ for both leptons as well as  the  $P_{NN}$ and $P_{TT}$ for the
 $\mu$ channel deviate slightly from each other. When the uncertainties of the form factors are considered, we detect considerable differences between full QCD and the HQET models predictions on behavior of 
 all double lepton polarization asymmetries with respect to the fourth family parameters.
\item Comparing to the other physical quantities, the double lepton polarization asymmetries are more sensitive to the mass of the fourth generation quark at lower values of $m_{t^\prime}$. This sensitivity
is large in HQET compared to the full QCD such that starting from the $m_{t^\prime}\simeq 200~GeV$, we see sizable deviations of the SM4 results with those of the SM in HQET approaximation. However, 
in the full theory
the discrepancy between the SM and SM4 results starts approaximately from $m_{t^\prime}\simeq 300~GeV$ and small compared to the HQET predictions.

\item When we consider only the central values of the form factors, except than the $P_{LN}$ and $P_{NT}$, the remaining double lepton polarization asymmetries
 grow increasing the $m_{t^\prime}$ and value of the $r_{sb}$. For $P_{LN}$ ($P_{NT}$), the maximum deviation belongs to the $r_{sb}=0.015$ and $m_{t^\prime}\simeq 450GeV$ ($r_{sb}=0.010$ and upper bound
of the $m_{t^\prime}$).
\end{itemize}
\newpage
\section{Conclusion}
We have performed a comprehensive analysis on the  $\Lambda_{b}\rightarrow \Lambda \ell^{+}\ell^{-}$ transition both in the SM4 and SM models. In particular, using the form factors entering the low energy
matrix elements both from full QCD as well as HQET, we have investigated  the branching ratio, forward-backward
asymmetry, double lepton polarization asymmetries and polarization
of the $\Lambda$ baryon.  We have observed that there are overall sizable differences between the predictions of the SM and SM4 on the considered physical quantities when
  $m_{t^\prime}\geq400~GeV$.  This can be considered as an indication of the existence of the fourth family quarks should we search for in the future experiments. The results also depicted overall  considerable
 differences between 
the predictions of the full QCD and those of the HQET. The orders of the branching ratios in both lepton channels show that  these decay channels can be detected at LHCb. Any measurement on the considered physical quantities and their comparison with the theoretical predictions can give valuable information
  about both
 nature of the participating baryons and existence of the fourth family quarks.

\end{document}